\documentclass[aps,prx,a4paper,twocolumn,superscriptaddress,longbibliography,10pt]{revtex4-2}

\usepackage{graphicx,bm,amssymb,physics,mathtools,mathrsfs, amsthm,color,xcolor,bbm,appendix}  
\usepackage{lineno}

\definecolor{magenta(dye)}{rgb}{0.79, 0.08, 0.48}
\definecolor{mediumred-violet}{rgb}{0.73, 0.2, 0.52}
\definecolor{neonfuchsia}{rgb}{1.0, 0.25, 0.39}
\definecolor{orange-red}{rgb}{1.0, 0.27, 0.0}
\definecolor{americanrose}{rgb}{1.0, 0.01, 0.24}
\definecolor{awesome}{rgb}{1.0, 0.13, 0.32}
\definecolor{blue}{rgb}{0.0, 0.0, 1.0}
\definecolor{cadmiumred}{rgb}{0.89, 0.0, 0.13}
\definecolor{candyapplered}{rgb}{1.0, 0.03, 0.0}
\definecolor{electricultramarine}{rgb}{0.25, 0.0, 1.0}
\definecolor{intblue}{rgb}{0.0, 0.18, 0.65}
\definecolor{navyblue}{rgb}{0.0, 0.0, 0.5}
\definecolor{jade}{rgb}{0.0, 0.66, 0.42}

\usepackage[colorlinks=true,linkcolor=jade,citecolor=intblue, urlcolor=cadmiumred, plainpages=false,pdfpagelabels]{hyperref}

\def\Tr{\operatorname{Tr}}

\newcommand{\hH}{\hat{H}}
\newcommand{\hq}{\hat{q}}

\newcommand{\hj}{J}
\newcommand{\dt}{\mathrm{d}\tau}
 \newcommand{\dom}{\mathrm{d}\omega}
\newcommand{\cost}{\cos\left(\lambda\tau\right)}
\newcommand{\sint}{\sin\left(\lambda\tau\right)}
\newcommand{\ct}{\cos\left(\frac{\theta}{2}\right)}
\newcommand{\st}{\sin\left(\frac{\theta}{2}\right)}
\newcommand{\ep}{e^{i\phi}}
\newcommand{\tu}{\tilde{u}}
\newcommand{\tv}{\tilde{v}}

\newcommand{\mc}[1]{\mathcal{#1}}

\newcommand{\io}{\mathfrak{i}}

\begin{document}

\title{Pointer States in the Born-Markov approximation}

\author{Uttam Singh}\email{uttam@cft.edu.pl}
\affiliation{Center for Theoretical Physics, Polish Academy of Sciences,\\ Aleja Lotnik\'ow 32/46, 02-668 Warsaw, Poland}
\author{Adam Sawicki}\email{a.sawicki@cft.edu.pl}
\affiliation{Center for Theoretical Physics, Polish Academy of Sciences,\\ Aleja Lotnik\'ow 32/46, 02-668 Warsaw, Poland}
\author{Jaros\l{}aw K. Korbicz}\email{jkorbicz@cft.edu.pl}
\affiliation{Center for Theoretical Physics, Polish Academy of Sciences,\\ Aleja Lotnik\'ow 32/46, 02-668 Warsaw, Poland}


\begin{abstract}
Explaining the emergence of classical properties of a quantum system through its interaction with the environment has been one of the promising ideas on how to understand the notorious quantum-to-classical transition. A pivotal role in this approach is played by, so called, pointer states which are quantum states least affected by the environment and are  ``carriers" of classical behavior.  We develop here a general method on how to find pointer states. Working within the Born-Markov approximation, we combine methods of group theory and open quantum systems to derive explicit equations describing pointer states. They contain variances squared of certain operators, thus resembling the defining equations of coherent states, but are in general different from the latter.  This shows that two notions of being ``the closest to the classical" -- one defined by the uncertainty relations and the other by the interaction with the environment -- are in general different. As an example, we study arbitrary spin-$J$ systems interacting with bosonic or spin thermal environments and find  explicitly pointer states for $J=1$.
\end{abstract}
\maketitle

\section{Introduction}
A comprehensive understanding  of how the classical reality emerges from the underlying quantum theory is one of the biggest and most fascinating challenges of modern physics. In this context, decoherence program \cite{Zeh1970, Joos2003, Zurek2003, Breuer2002, Schlosshauer2007} has been highly successful in explaining the loss of quantum properties through the interaction with uncontrolled degrees of freedom (environment); see e.g. \cite{Chiorescu2003, Deleglise2008, Haffner2008, Hornberger2012, Moser2014, Tighineanu2018, Fein2019}.  The inevitable, in most realistic situations, interactions with the environment lead to delocalization and destruction of phase relations, making certain quantum superpositions unobservable. On the other hand, the same process distinguishes some preferred states \cite{Zurek1981, Zurek1982}, which are least affected, and thus the system is most likely to be found in one of these states. In this sense, the perceived classicality can be explained through properties of certain robust quantum states \cite{Joos2003}. 

Determining the preferred states, known as pointer states, in the general case has been a difficult open task since their introduction in \cite{Zurek1981}. Several formal definitions were given with the most fruitful being the predictability sieve idea \cite{Zurek1993b}, defining pointer states as states producing least entropy (for others see e.g. \cite{Diosi2000}). Various examples of pointer states have been found so far, with the the best known \cite{Zurek1993} in the Quantum Brownian Motion (QBM)  model \cite{Feynman1963, Ullersma1966, Joos2003, Schlosshauer2007}, where they happen to be the Glauber-Sudarshan coherent states \cite{Glauber1963, Sudarshan1963}. Minimum uncertainty states  were also proven to be universal pointer states for a general, linearly coupled free open evolution and that decoherence to them is generic \cite{Eisert2004}, which is an extension of an earlier result \cite{Diosi2000}, obtained in a simpler Markovian model. A general question when generalized coherent states \cite{Perelomov1986} are the preferred states was analyzed in \cite{Boixo2007}, using the Gorini-Kossakowski-Lindblad-Sudarshan (GKLS) \cite{Kossakowski1972, Davies1974, Gorini1976, Lindblad1976} master equation and group-theoretical methods, showing that it is not generically the case unless specific conditions are met. Other results include \cite{Paraoanu1998}, where similar findings to \cite{Zurek1993} were derived in the GKLS formalism, \cite{Khodjasteh2011}, where the problem of how to engineer the coupling to turn a given pure state into a pointer state was studied in a non-Markovian setup, and \cite{Donker2017} where pointer states for anti-ferromagnetic systems were numerically analyzed. 

In this work we derive a general framework for finding  pointer states in the Born-Markov approximation with a linear coupling. We do so under a broad assumption of an existence of  some Lie group structure behind the dynamics, which covers, among others, the canonical models of decoherence \cite{Schlosshauer2007}. We analyze in detail the case of a compact, semi-simple group, but our method applies to other groups too as we show on the QBM example of \cite{Zurek1993}. The resulting conditions take form of an optimization problem involving sum of variances of certain operators and are in general different from the ones defining coherent states for the group, confirming that open dynamics selects its own robust states, different from the static minimal uncertainty principle (cf. \cite{Boixo2007}). As a concrete example, we apply our method to thermal spin models with arbitrary spin-$J$ central systems interacting with either bosonic or spin environments. For $J=1$ we explicitly find the pointer states. 

The use of the Born-Markov rather than the GKLS approach  can be criticized on the ground that it lacks complete-positivity, which has been a matter of an ongoing debate (see e.g. \cite{Winczewski2021, Tupkary2022} and the references therein), despite the immense predictive power of the former \cite{Schlosshauer2007}.  We use the Born-Markov equation as it allows for a more direct connection with the underlying microscopic model by avoiding the secular approximation \cite{Breuer2002} and the associated problems \cite{Tupkary2022}.  We stress that is not our aim to add to the debate but rather to exploit the potential of the Born-Markov equation in a search for pointer states.

The paper is organized is as follows: In Sec. \ref{sec:pred-sieve}, we start with a general Born-Markov master equation and introduce the framework to incorporate a group structure on the free and the interacting Hamiltonians. Then based on this group structure, we derive our main result, namely, the conditions giving rise to the approximate pointer states. In Sec. \ref{sec:QBM}, we apply our framework to the example of QBM and discuss why the coherent states are the approximate pointer states in this particular case. In Sec. \ref{sec:spin-1}, we consider the example of spin-$1$ particles and show explicitly that the coherent states are not the approximate pointer states. Finally, we conclude in Sec. \ref{sec:conclusion} with a discussion on ramifications of our results along with the future directions. In Appendix \ref{append:spin-1}, we provide detailed calculations for spin-$1$ example in various regions of temperature of the environment.

\section{Pointer states and group-theoretical methods  in the Born-Markov approximation.}
\label{sec:pred-sieve}
We consider an open system model, $H=H_0+H_E+H_I$, where a system of interest $S$, governed by the free Hamiltonian $H_0$, is coupled  to the environment $E$ via a bilinear interaction term $H_{I}=A\otimes \mathcal E$. The most general coupling is a sum of such terms \cite{Schlosshauer2007} and a generalization of our method to such couplings is straightforward.
We will assume the weak-coupling limit and that the conditions of the Born-Markov approximation hold. Then system's reduced density matrix $\rho(t)$ satisfies the Born-Markov master equation \cite{Redfield1957, Bloch1957, Schlosshauer2007}:
\begin{align}
\label{eq:mast-eq}
\dot\rho(t)&= -\io [H_0, \rho(t)] -\int_0^{\infty} \dt \nu(\tau)\left[A,\left[A(-\tau),\rho(t)\right]\right]\nonumber\\
&+\io \int_0^{\infty} \dt \eta(\tau)\left[A,\left\{A(-\tau),\rho(t)\right\}\right],
\end{align}
where $\{A,B\}=AB+BA$, 
\begin{align}\label{free_ev}
A(-\tau)=e^{-\io H_0\tau}Ae^{\io H_0\tau},
\end{align} 
and $\nu(\tau)$, $\eta(\tau)$ are respectively the the noise and dissipation kernels defined via the environment correlation function:
\begin{align}\label{nu}
\tr_E\left[\rho_E(0) \mathcal E(\tau) \mathcal E\right] \equiv \nu(\tau)-\io\eta(\tau), 
\end{align}
$\mathcal E(\tau) =e^{\io H_E\tau}\mathcal E e^{-\io H_E\tau}$. The environment state $\rho_E$ is arbitrary here as long as it is stationary $[H_E, \rho_E]=0$, which is the standard assumption \cite{Joos2003, Breuer2002, Schlosshauer2007}. 

The predictability sieve looks for those states of the system, which generate the least entropy during the evolution \eqref{eq:mast-eq}.  A convenient measure is the linear entropy (c.f. \cite{Diosi2000, Zurek2003}):
\begin{align}\label{entropy}
s(\rho)=1-\Tr[\rho^2],
\end{align}
connected to the loss of purity. We thus assume a pure initial state:
\begin{align}
\rho(0)=\ket\psi\bra\psi 
\end{align}
and ask how much of the purity is lost during the evolution \cite{Zurek1993}. 
Let us now assume  that there exists a Lie group $G$ of dimension $N$, such that $H_0$ and $A$ are from the Lie algebra of $G$ and in particular can be represented using generators $\{X_i\}_{i=1}^N$ of $G$: 
\begin{align}
& H_0 \equiv X_N\label{H0},\\
& A \equiv \sum_{j=1}^N a_j X_j, \label{A}
\end{align}
(we choose $H_0$ as one of the generators and leave $A$ arbitrary only for the sake of definiteness). 
This is somewhat similar to the approach of \cite{Boixo2007}, however instead of imposing the secular approximation and  studying the algebra generated by the GKLS jump operators and the associated generalized coherent states, we assume the existence of a dynamical group already at the microscopic level.
Both $H_0$ and $A$ must be Hermitian for obvious physical reasons, which motivates the assumption that the Lie algebra in question is real and spanned by Hermitian operators, $X_j^\dagger=X_j$. Further assuming  they can be represented in finite dimension, $G$ becomes a subgroup of sufficiently large unitary group and thus compact. 
In light of these identifications, the free evolution, given by Eq. \eqref{free_ev}, becomes the adjoint action of $G$ in its Lie algebra. In particular, we have
\begin{align}\label{Ad}
    X_j(-\tau)  =e^{-\io X_N\tau}X_j e^{\io X_N\tau}=\sum_{k} R^{N}_{\phantom{i}jk}(-\tau) X_k,
\end{align}
where $R^{N}_{\phantom{i}jk}(0)=\delta_{jk}$. The indices $i,j,k,l\dots \in 1\dots N $ are the Lie algebra indices of $G$. Matrix $R^N$ is nothing but the exponent of the structure constants $f_{ijk}$ of $G$, arranged into $N\times N$ matrix (the matrix of the $\text{ad}_{X_N}$ action) and is given by 
\begin{align}
R^{N}_{\phantom{i}jk}(t)=\left[ e^{\io t ~\text{ad}_{X_N}}\right]_{jk},\ [\text{ad}_{X_N}]_{jk}=\io f_{Njk},\label{adN}
\end{align}
where $f_{ijk}$ are defined via $[X_i, X_j]=\io \sum_{k}f_{ijk}X_k$. In what follows we will omit the index $N$ for simplicity, writing $R_{jk}(\tau)$. Substituting Eq. \eqref{Ad} into Eq. \eqref{eq:mast-eq}, we obtain
\begin{align}
\dot\rho(t) 
&= -\io [X_N, \rho(t)]-\sum_{jkl} a_l a_j D_{jk} \left[X_l,\left[X_k,\rho(t)\right]\right]\nonumber\\
&+\io \sum_{jkl}a_l a_j\gamma_{jk}\left[X_l,\left\{X_k,\rho(t)\right\}\right],
\end{align}
where we have introduced constants (assuming the correlation function \eqref{nu} is regular enough for the integrals to exist):
\begin{align}\label{const}
&D_{jk}=\int_0^{\infty} \dt \nu(\tau) R_{jk}(-\tau),\ \gamma_{jk}= \int_0^{\infty} \dt \eta(\tau) R_{jk}(-\tau).
\end{align}
We can now express the change of the entropy \eqref{entropy} as
\begin{align}
\frac{1}{2}\dot s
&= \sum_{jkl} a_j a_l D_{jk} \left(\Tr\left[\rho^2\{X_l, X_k\}\right]-2\Tr\left[\rho X_l\rho X_k \right]\right)\nonumber\\
& + \sum_{jklm}a_j a_l\gamma_{jk}f_{lkm}\Tr\left[\rho^2X_m\right],
\end{align}
To make the further analysis feasible, we may assume as a first approximation \cite{Zurek1993} that the state $\rho(t)$ at the right hand side above is approximately pure and evolves 
according to the free evolution, i.e.,
\begin{align}
\rho(t)\approx e^{-\io X_N t} \ket\psi\bra\psi e^{\io X_N t}.
\end{align}
Using Eq. \eqref{Ad} we obtain
\begin{align}
\frac{1}{2}\dot  s \approx &\sum_{jklmn} a_j a_l D_{jk} R_{lm}(t) R_{kn}(t) C_{mn}+\label{dts}\\
& \sum_{jklmn} a_j a_l \gamma_{jk}f_{lkm} R_{mn}(t) \langle X_n\rangle, \label{dts2}
\end{align}
where
\begin{align}\label{Cmn}
C_{mn}=\langle \{X_m, X_n\}\rangle -2\langle X_m\rangle \langle X_n \rangle
\end{align}
is the  covariance matrix calculated in the initial state and $\langle X_m\rangle=\langle \psi|X_m|\psi\rangle$ is the average in the initial state.

To analyze Eqs. \eqref{dts} and \eqref{dts2} further, we look closer at the structure of matrices $R_{jk}(t)$. For compact group $G$, there is a Killing-Cartan form, $h_{jk}=\Tr[\text{ad}_{X_j}\text{ad}_{X_k}]$ (see e.g. \cite{kirillov2008}) on the Lie algebra of $G$ with a definite signature and which serves as a metric. Moreover, $h_{jk}$ is preserved by the adjoint action \eqref{Ad}. For simplicity, we will also assume that this metric is non-degenerate (so the group is semi-simple), but this is not crucial. We can then diagonalize the metric and assume $h_{jk}\propto \delta_{jk}$ so that the matrices $R_{jk}(t)$ of the adjoint action become orthogonal matrices. Since they are in the connected component of the unity by taking $t\to 0$ in Eq. \eqref{Ad}, they are also special orthogonal, i.e. $R(t)\in SO(N)$. Furthermore, from Eq. \eqref{Ad} we obviously have  $e^{\io X_N t}X_N e^{-\io X_Nt} = X_N$, i.e. the N-th row has just one element $R_{Nk}(t) =\delta_{Nk}$. From the orthogonality $R(t)^TR(t)={\bf 1}$, the $N$-th column has the same property and  $R(t)$ is of the form:
\begin{align}\label{axis}
R(t)=\left[\begin{matrix} SO(N-1) & 0\\ 0 & 1 \end{matrix}\right],
\end{align}
i.e. it is a rotation around the axis of $X_N$ as one could easily guess from Eq. \eqref{Ad}. In most physical models the non-dynamical part $a_N X_N$ of $A$ is usually neglected but we keep it here for completeness.
The $SO(N-1)$ part can be further decomposed into a direct sum of $2D$ rotations by a change of basis in the Lie algebra. For odd $N-1$, i.e. even $N$, one of the blocks will again be  $1$. Summarizing, we can write the following decomposition of the full matrix $R(t)$:
\begin{align}\label{canon}
R(t)=O^T\left[\oplus_\alpha R_\alpha(t)\right] O,\ R_\alpha(t)=\left[\begin{matrix} \cos t \Omega_\alpha & \sin t \Omega_\alpha \\ -\sin t \Omega_\alpha  & \cos t\Omega_\alpha \end{matrix}\right],
\end{align}
where $O\in SO(N)$ (we prefer to use the full dimensional matrices extended by $1$ on the diagonal for the ease of the index notation later) and the last block $R_{\alpha_N}(t)=1$. Additionally, for even $N$ the one before the last block is also trivial, i.e., we have
\begin{align}\label{canon2}
OR(t)O^T=\left[\begin{matrix} R_1(t) &  &  &  &  & \\ & \cdot & & & & \\ & & \cdot & & &\\ & & & \cdot & &\\ & & & & \cdot & \\& &  & & & 1\\
\end{matrix}\right],\ \left[\begin{matrix} R_1(t) &  &  &  &  & \\ & \cdot & & & & \\ & & \cdot & & &\\ & & & \cdot & &\\ & & & & 1 & \\& &  & & & 1\\
\end{matrix}\right]
\end{align}
for odd and even $N$, respectively. The form of the time dependence of rotations $R_\alpha(t)$ comes from the antisymmetry of the generator of the adjoint representation, Eq. \eqref{Ad}, which in turn follows from the differential form of the invariance of the Cartan-Killing form:
\begin{align}\label{fantisym}
f_{ijk}=-f_{ikj},
\end{align}
making $f_{ijk}$ totaly anti-symmetric as by definition $f_{ijk}=-f_{jik}$.

Summarizing, for each particular model, the l.h.s. of Eq. \eqref{dts} is a finite linear combination of trigonometric functions and their squares, which, in principle, can be  integrated (cf. \cite{Zurek1993}). However, the instantaneous value of the entropy production $s(t)$ may not be the most indicative quantity due to its time fluctuations. Here we choose its long-time average as more representative:
\begin{align}
\bar s =\lim_{\tau\to\infty}\frac{1}{\tau}\left[s(\tau)-s(0)\right]= \lim_{\tau\to\infty}\frac{1}{\tau} \int_0^\tau dt \dot s(t),
\end{align}
meaning we average the entropy production over times much longer than any other time scales. We apply the above time-averaging to Eq. \eqref{dts} using Eq. \eqref{canon}. The first term gives
\begin{align}
&\overline{R_{lm}(t) R_{kn}(t)} = \nonumber \\
& = \sum_{k'l'm'n'} O_{l'l}O_{k'k} 
\sum_{\alpha\beta}\overline{R^{(\alpha)}_{l'm'}(t) R^{(\beta)}_{k'n'}(t)}O_{m'm}O_{n'n}\nonumber\\
& = \sum O_{l'l}O_{k'k} 
\sum_{\alpha}\frac{1}{2}\left[\delta^{(\alpha)}_{l'm'}\delta^{(\alpha)}_{k'n'}+\epsilon^{(\alpha)}_{l'm'}\epsilon^{(\alpha)}_{k'n'}\right]O_{m'm}O_{n'n}, \label{RR1}
\end{align}
where $R^{(\alpha)}_{jm}$ are the matrices of $R_\alpha$ embedded naturally into the whole space by adding rows and columns of zeros, i.e. $R^{(\alpha)}_{jm} =0$ when $jm$ are outside of the $\alpha$-subspace. We have also used the fact that 
\begin{align}\label{RR2}
\overline{R^{(\alpha)}_{lm}(t) R^{(\beta)}_{kn}(t)}=\delta_{\alpha\beta}
\frac{1}{2}\left[\delta^{(\alpha)}_{lm}\delta^{(\alpha)}_{kn}+\epsilon^{(\alpha)}_{lm}\epsilon^{(\alpha)}_{kn}\right],
\end{align}
for non-trivial blocks. Trivial blocks give simply $1$. Here $\epsilon_{lm}$ is the totally anti-symmetric Levi-Civita symbol in dimension two. Eq. \eqref{RR2} follows from a direct entry-by-entry averaging of $R_{\alpha}(t)\otimes R_{\beta}(t)$, using the explicit form, Eq. \eqref{canon}, where only the quadratic terms with the same angular velocity, i.e., $\sin^2t\Omega_\alpha $ and $\cos^2t\Omega_\alpha$, give non-zero contributions (We assume a generic situation with no special relations between eigenfrequencies $\Omega_\alpha$ for different $\alpha$). Let us define rotated generators as
\begin{align}
\tilde X_{m'} \equiv \sum_{m} O_{m'm} X_{m}\label{rotated}
\end{align}
 and the corresponding covariance matrix $\tilde C_{mn}$. We recall that $\tilde X_N=X_N$ by the construction of $O$, cf. Eq. \eqref{axis}. We then have
\begin{align}
& \sum_{mnm'n'}\left[\delta^{(\alpha)}_{lm}\delta^{(\alpha)}_{kn}+\epsilon^{(\alpha)}_{lm}\epsilon^{(\alpha)}_{kn}\right] O_{mm'}O_{nn'} C_{m'n'}\nonumber\\
&=\left[\tilde C_\alpha + \epsilon \tilde C_\alpha \epsilon^T\right]_{lk}=\Tr[\tilde C_\alpha] \delta^{(\alpha)}_{lk},\label{trC}
\end{align}
where $\tilde C_\alpha$ is the projection of $\tilde C$ on the $\alpha$-subspace and in the last step we used the symmetry of  $\tilde C_\alpha$ (inherited from $\tilde C$). Using this, we can write
\begin{align}
& \sum_{jkl} a_j a_l D_{jk} O_{l'l}O_{k'k} = \sum_\alpha \sum_{i'jkm'} \tilde a_{l'} a_j O_{i'j} D^{(\alpha)}_{i'm'} O_{m'k} O_{k'k}\nonumber\\
& =  \sum_\alpha \sum_{i} \tilde a_{l'}\tilde a_{i'} D^{(\alpha)}_{i'k'},\label{aa}
\end{align}
where we introduced a rotated vector $\tilde a_i \equiv \sum_j O_{ij} a_j$ and decomposed the matrix $D_{jk}$ by inserting decomposition, given by Eq. \eqref{canon} into Eq. \eqref{const} and defined
\begin{align} \label{Dbeta}
 D^{(\alpha)}_{jk} \equiv \int_0^{\infty} \dt \nu(\tau) R^{(\alpha)}_{jk}(-\tau) = \left[\begin{matrix} D_\alpha & f_\alpha \Omega_\alpha \\ -f_\alpha \Omega_\alpha & D_\alpha \end{matrix} \right]_{jk}.
\end{align}
The coefficients $D_\alpha$ and $f_\alpha$ are generalized normal and anomalous diffusion coefficients, corresponding to the eigenfrequency $\Omega_\alpha$:
\begin{align}
& D_\alpha =   \int_0^{\infty} \dt \nu(\tau)\cos\tau\Omega_\alpha
\label{Dalpha}\\
& f_\alpha =   -\frac{1}{\Omega_\alpha}\int_0^{\infty} \dt \nu(\tau)\sin\tau\Omega_\alpha
\end{align}
We now substitute Eqs. \eqref{RR1}, \eqref{trC}, and \eqref{aa} into Eq. \eqref{dts} to finally obtain
\begin{align}
& \sum_{jklmn} a_j a_l D_{jk} \overline{R_{lm}(t) R_{kn}(t)} C_{mn}\nonumber\\
&= \frac{1}{2}\sum_{\alpha\beta}\sum_{i'l'k'} \tilde a_{i'} \tilde a_{l'} D^{(\beta)}_{i'k'} \Tr[\tilde C_\alpha] \delta^{(\alpha)}_{l'k'} \nonumber\\
&=\frac{1}{2}\sum_\alpha ||{\bf \tilde a}_\alpha||^2 D_\alpha \Tr[\tilde C_\alpha]  \nonumber\\
& =\sum_{\alpha<\alpha_N} ||{\bf \tilde a}_\alpha||^2 D_\alpha \left[\Delta \tilde X_{\alpha 0}^2  + \Delta \tilde X_{\alpha 1}^2 \right]+ a_N^2 D_0 \Delta X_{N}^2\label{I},
\end{align}
and for even $N$ there is an additional term $\tilde a_{N-1}^2 D_0 \Delta \tilde X_{N-1}^2$. Here $\alpha_N=\left \lceil{(N-1)/2}\right \rceil$,  $||{\bf \tilde a}_\alpha||^2$ is the norm squared of the projection of the vector  ${\bf \tilde a}$ on the $\alpha$-subspace. We used the fact that matrices ${\bf D}_\beta$ and $\mathbf 1_\alpha$ are supported in different subspaces for $\alpha\ne \beta$, and the explicit form, Eq. \eqref{Dbeta}, to calculate the quadratic form. Further, in the last step we used the definition of $\tilde C_\alpha$, given by Eq. \eqref{Cmn}, but with the rotated generators (Eq. \eqref{rotated}). Moreover, $\tilde X_{\alpha 0}$, $\tilde X_{\alpha 1}$ denote the two generators in the $\alpha$-subspace and $\Delta \tilde X_{\alpha i}^2 = \langle \tilde X_{\alpha i}^2\rangle - \langle \tilde X_{\alpha i}\rangle^2 $ are their variances in the initial state $\ket\psi$.  Finally, we introduced
$D_0=\int\dt \nu(\tau)$, provided the integral exists (e.g. for Ohmic and super-Ohmic bosonic environments). 

In the similar fashion we now analyze the term, Eq. \eqref{dts2}, linear in $R_{jk}(t)$. Let us first assume odd group dimension $N$. It is then clear from Eqs. \eqref{axis} and \eqref{canon} that the only term that survives the time averaging is $\overline{R_{NN}(t)}=R_{NN}(t)=1$, the rest being zero, so that
\begin{align}
\sum_{jklmn} a_j a_l \gamma_{jk}f_{lkm} \overline{R_{mn}(t)} \langle X_n\rangle =- \sum_{jkl} a_j a_l\gamma_{jk}f_{Nkl} \langle X_N\rangle,\label{linear1}
\end{align}
where we have used the total antisymmetry of $f_{ijk}$, cf. Eq. \eqref{fantisym}. Let us now decompose matrices $\gamma_{jk}$ and $f_{Njk}$ following the decomposition, Eq. \eqref{canon}. Substituting Eq. \eqref{canon} into Eq. \eqref{const} we obtain
\begin{align}
\bm\gamma = \sum_{\alpha} O^T \bm{\gamma}_{\alpha} O,\ \bm{\gamma}_{\alpha} = \left[\begin{matrix} -\tilde\Omega^2_\alpha & -\gamma_\alpha \Omega_\alpha \\ \gamma_\alpha \Omega_\alpha & -\tilde\Omega^2_\alpha \end{matrix} \right],
\end{align}
where 
\begin{align}
& \tilde\Omega_\alpha^2=- \int_0^{\infty} \dt \eta(\tau)\cos\tau\Omega_\alpha,\\
& \gamma_\alpha =   \frac{1}{\Omega_\alpha}\int_0^{\infty} \dt \eta(\tau)\sin\tau\Omega_\alpha \label{gammaalpha}
\end{align}
are generalized frequency shift and the momentum damping coefficients, corresponding to the eigenfrequency $\Omega_\alpha$ (cf.  \cite{Schlosshauer2007}). In the similar way, Eq. \eqref{canon} implies via the differentiation of Eq. \eqref{adN}, the block-diagonal form of the $\text{ad}_{X_N}$ matrix as
\begin{align}
\text{ad}_{X_N} = \sum_{\alpha<\alpha_N} O^T c_{\alpha} O,\ c_{\alpha} = \left[\begin{matrix} 0 & -\io \Omega_\alpha \\ \io \Omega_\alpha & 0 \end{matrix} \right],
\end{align}
where the last block is zero due to Eq. \eqref{axis}. Recalling that $[\text{ad}_{X_N}]_{jk}=\io f_{Njk}$; using matrix/vector notation and the natural embeddings of $\gamma_\alpha$ and $c_\alpha$ into the whole space, we obtain from Eq. \eqref{linear1} that
\begin{align}
 \sum_{jkl} a_j a_l \gamma_{jk}f_{Nkl} &=-\io \langle {\bf a} | \bm\gamma\cdot \text{ad}_{X_N} {\bf a}\rangle\nonumber\\
& =-\io  \sum_{\alpha\beta} \langle {\bf a} | O^T\bm\gamma_\alpha OO^T c_\beta O {\bf a}\rangle\nonumber\\
& = -\io
\sum_{\alpha<\alpha_N} \langle {\bf \tilde a} |\bm\gamma_\alpha\cdot c_\alpha {\bf \tilde a}\rangle\nonumber \\
& =- \sum_{\alpha<\alpha_N} ||{\bf \tilde a}_\alpha||^2 \Omega_\alpha^2 \gamma_\alpha ,
\end{align}
where we used the explicit forms of $\gamma_\alpha$ and $c_\beta$ and the fact that they are supported in different subspaces for $\alpha\ne \beta$. Thus, the linear term, Eq. \eqref{linear1}, is equal to
\begin{align}
 \sum_{\alpha<\alpha_N} ||{\bf \tilde a}_\alpha||^2 \Omega_\alpha^2 \gamma_\alpha \langle X_N\rangle.  \label{linear2}
\end{align}

For even dimension $N$, the situation is more complicated. There is one more non-zero element in $\overline{R_{mn}(t)}$, corresponding to the $(N-1,N-1)$ element of the canonical form of $R(t)$: $[O R(t) O^T]_{N-1,N-1}=1$, cf. Eq. \eqref{canon2}.
It leads to an additional term in Eq. \eqref{linear1}:
\begin{align}
& \sum_{jklmn} a_j a_l \gamma_{jk}f_{lkm} O_{N-1,m} O_{N-1,n} \langle X_n\rangle\nonumber\\
&= -\sum_{jklm} a_j a_l \gamma_{jk} \left(\sum_m O_{N-1,m}f_{mkl}\right) \langle \tilde X_{N-1}\rangle\nonumber\\
&= \io \langle {\bf a} | \bm\gamma\cdot \text{ad}_{\tilde X_{N-1}} {\bf a}\rangle \langle \tilde X_{N-1}\rangle\nonumber\\
&= \io \sum_\alpha\langle {\bf \tilde a} |\bm\gamma_\alpha \cdot O\text{ad}_{\tilde X_{N-1}} O^T{\bf \tilde a}\rangle \langle \tilde X_{N-1}\rangle\nonumber\\
&= \io \sum_\alpha\langle {\bf \tilde a} |\bm\gamma_\alpha \cdot \widetilde{\text{ad}}_{\tilde X_{N-1}}{\bf \tilde a}\rangle \langle \tilde X_{N-1}\rangle,
\end{align}
where $\text{ad}_{\tilde X_m}$ is the $\text{ad}$ operator of the transformed generator $\tilde X_m = \sum O_{mm'}X_{m'}$ and $\widetilde{\text{ad}}_{\tilde X_{N-1}}=O\text{ad}_{\tilde X_{N-1}}O^T$ is the matrix of 
$\text{ad}_{\tilde X_{N-1}}$ transformed to the new basis so that $[\tilde X_i, \tilde X_j]=\io \sum_{k} [\widetilde{\text{ad}}_{\tilde X_i}]_{jk} \tilde X_k$. This is as much as can be said in a general case.

Summarizing, for an odd dimension $N$, the asymptotic entropy production reads
\begin{align}
&\frac{1}{2} \overline s\approx \sum_{\alpha<\alpha_N} ||{\bf \tilde a}_\alpha||^2 D_\alpha \left[\Delta \tilde X_{\alpha 0}^2  +\Delta \tilde X_{\alpha 1}^2 +\frac{\Omega_\alpha^2 \gamma_\alpha}{D_\alpha} \langle X_N\rangle\right]\label{Noddfinal1}\\
&+ a_N^2 D_0 \Delta X_{N}^2\label{Noddfinal2}
\end{align}
For even $N$ it reads
\begin{align}
&\frac{1}{2}\overline s\approx \sum_{\alpha<\alpha_N-1} ||{\bf \tilde a}_\alpha||^2 D_\alpha \left[\Delta \tilde X_{\alpha 0}^2  + \Delta \tilde X_{\alpha 1}^2 +\frac{\Omega_\alpha^2 \gamma_\alpha}{D_\alpha} \langle X_N\rangle\right]\nonumber\\
&+ \tilde a_{N-1}^2 D_0 \Delta \tilde X_{N-1}^2+a_N^2 D_0 \Delta X_{N}^2\nonumber\\
&+\io \sum_\alpha\langle {\bf \tilde a} |\bm\gamma_\alpha \cdot \widetilde{\text{ad}}_{\tilde X_{N-1}}{\bf \tilde a}\rangle \langle \tilde X_{N-1}\rangle. \label{Nevenfinal3}
\end{align}
To simplify the above expressions further, let us consider a low-damping limit $\gamma_{jk}/D_\alpha \approx 0$, which holds e.g. in high-temperature environments  \cite{Schlosshauer2007, Breuer2002},  
and drop the non-dynamical term $a_N X_N$ from  Eq. \eqref{A} as it is usually the case. This gives:
\begin{align}
\frac{1}{2}  \overline s\approx \sum_{jk} g_{jk} \left[\langle \tilde X_j \tilde X_k\rangle - \langle\tilde X_j\rangle\langle\tilde X_k\rangle\right],\label{highT}
\end{align}
where we introduced an environment-dependent metric
\begin{align}\label{g}
g_{jk}=\left[\begin{matrix} ||{\bf \tilde a}_{\alpha_1}||^2 D_{\alpha_1} \bm 1_2 &  &  &  &  & \\ & \cdot & & & & \\ & & \cdot & & &\\ & & & \cdot & &\\ & & & & \cdot & \\& &  & & & 0\\
\end{matrix}\right],
\end{align}
where for even $N$ the last non-zero block is equal to $\tilde a_{N-1}^2 D_0$. There is a formal similarity between  \eqref{highT} and  the definition of the generalized coherent states as the states that minimize the $G$-invariant dispersion $\langle \Delta C\rangle =\sum h_{jk}[\langle \tilde X_j \tilde X_k\rangle - \langle\tilde X_j\rangle\langle\tilde X_k\rangle]$  \cite{Perelomov1986}. 
However, the metric $g_{jk}$ is in general different from the Killing-Cartan form $h_{jk}$, which here is $\propto \bm 1$. For example, in thermal models $g_{jk}$ can non-trivially depend on the temperature via the diffusion coefficients $D_\alpha$.  Thus
the states minimizing Eq. \eqref{highT} are in general different from the coherent states for $G$ unless $g_{jk}\propto h_{jk}$ or more generally $\overline s \propto \langle \Delta C\rangle$ \cite{Boixo2007}.

Although we have assumed for definiteness' sake a compact, semi-simple dynamical group $G$, our method is not restricted to such groups only. We used the fact that  $R^{N}_{\phantom{i}jk}(t)$, \eqref{adN}, is from a group with a known canonical form \eqref{canon}.  It is clear that other groups with known canonical forms, e.g. symplectic,  can be analyzed by the above method. It can also happen that a non-compact group leads to orthogonal $R^{N}_{\phantom{i}jk}(t)$ as we show below. 

\section{ Quantum Brownian Motion case}
\label{sec:QBM}
It is interesting to revisit from the current perspective the seminal result of \cite{Zurek1993}, which shows that for the  quantum Brownian motion model the pointer states are the coherent states. The model is described by
$H_0=P^2/2M+M\Omega^2 Q^2/2$ and $A=Q$. The group $G$ that can be associated with the dynamics is generated by the operators $\bm X =\{Q, P, \bm 1, H_0\}$ and is known as the oscillator group \cite{Streater1967}. The group is non-compact
(it is a projective representation of $Sp(2,\bm R)$) but after the rescaling $q=\sqrt M\Omega Q$, $p=(1/\sqrt M) P$, the matrix of the adjoint action (see Eq. \eqref{Ad}) generated by $h_0=(q^2+p^2)/2$ becomes orthogonal (with the Lie algebra basis $\{q,p,\bm 1, h_0\}$ and $[q,p]=\io \Omega$). In particular,
\begin{align}
R(t)=\left[\begin{matrix} \cos\Omega t & \sin\Omega t &  & \\ -\sin\Omega t & \cos\Omega t &  & \\ & & 1 & \\ & & & 1 \\ \end{matrix}\right].
\end{align}
The matrix $R(t)$ is already in the canonical form, Eq. \eqref{canon2}. We can thus use our procedure and obtain the high temperature entropy production equation (Eq. \eqref{highT} with $a_1=1$ and the rest $a_i$'s zero):
\begin{align}
\overline s\approx 2D\left[\Delta q^2+\Delta p^2\right]=2DM\Omega^2\left[\Delta Q^2+\frac{\Delta P^2}{M^2\Omega^2}\right],\label{QBM}
\end{align}
Modulo an unimportant prefactor, the above equation is the same as obtained in Ref. \cite{Zurek1993} via a direct calculation.  $D$ is given by Eq. \eqref{Dalpha} with $\Omega_{\alpha}=\Omega$. It now so happens that the r.h.s. of Eq. \eqref{QBM} corresponds to an invariant dispersion for a subgroup $H$ of $G$--the Heisenberg-Weyl group  generated by $\{Q, P, \bm 1\}$. The minimization of Eq. \eqref{QBM} leads to the coherent states of $H$, which are the celebrated Glauber-Sudarshan coherent states $\ket\alpha$ \cite{Sudarshan1963, Glauber1963}, and which are also coherent states for $G$ \cite{Perelomov1986}. This situation, however, is rather exceptional as e.g. it is well known  that taking higher order than quadratic polynomials in $Q$, $P$ will not lead to any group structures, which in turn is connected to the problems of canonical quantization \cite{Groenewold1946}. 

\section{Spin-$J$ systems}
\label{sec:spin-1}
\begin{figure}
\centering
\includegraphics[width=80mm]{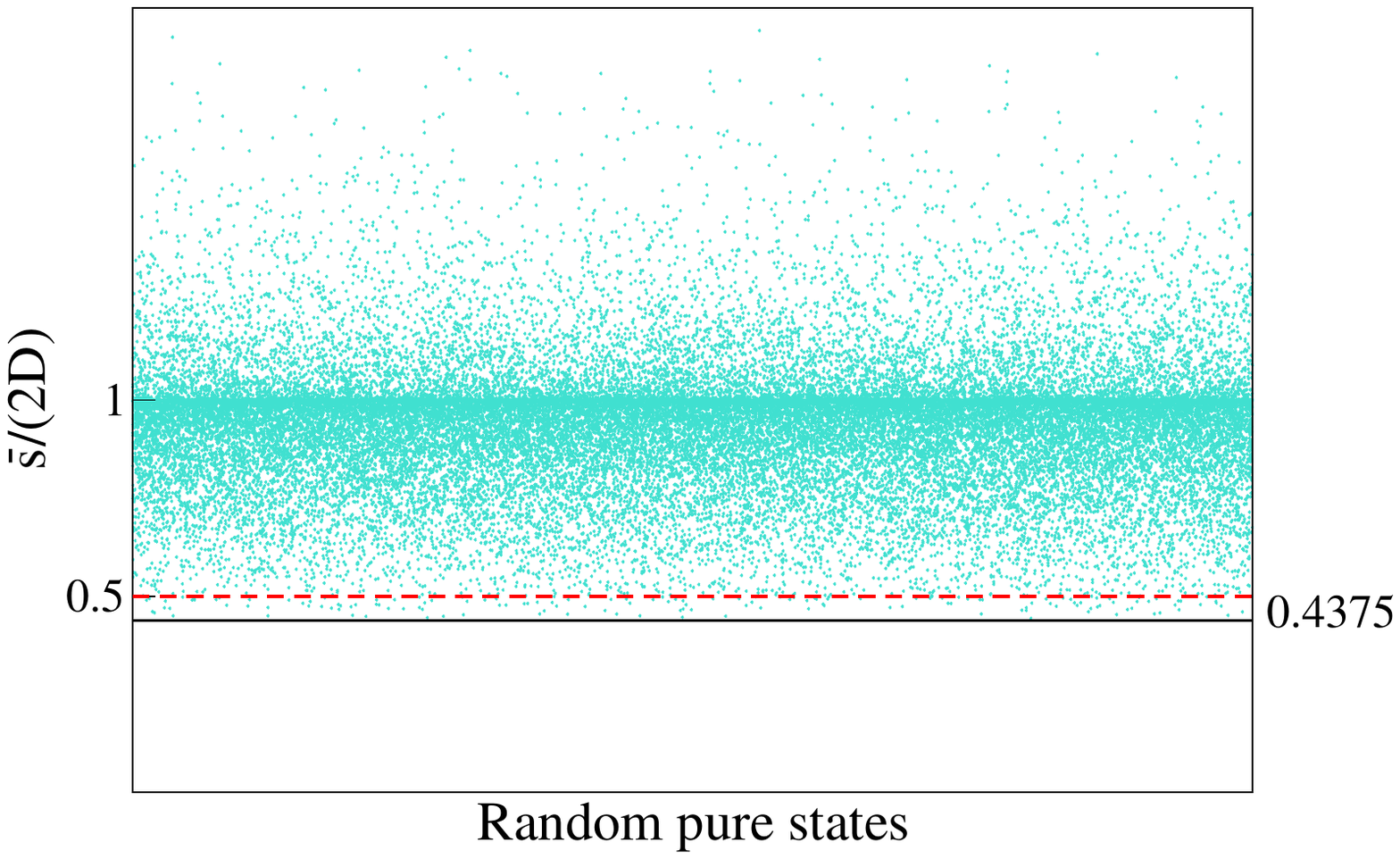}
\caption{Values of the rescaled entropy production $\overline{s}/2D$ for random pure states in the high temperature limit (see Eq. \eqref{spinJht}) for spin-$1$ system. Each blue dot in the plot corresponds to the value of the rescaled entropy production for a single random pure state. The red dashed horizontal line corresponds to the minimum value of the rescaled entropy production $\overline{s}/2D$ obtained by using spin coherent states. The black horizontal line indicates the true minimum value, $0.4375$, and it is strictly lower than the value obtained for the spin-coherent states.}
\label{fig:scatter-plot}
\end{figure}

As a further illustration, we consider a class of models where a central spin-$J$ interacts with a thermal environment.
As our general method is quite insensitive to the type of the environment, as long as the auto-correlation function is sufficiently regular, it will not matter below if the 
environment is bosonic, with the total Hamiltonian:  
\begin{align}\label{Hsb}
H=\Omega \hj_z+\sum_{i}\omega_i a_i^\dagger a_i - \hj_x \sum_{i}(g_i a_i^\dagger + g_i^* a_i),
\end{align}
or spin:
\begin{align}\label{Hss}
H=\Omega \hj_z+\sum_{i}\frac{\omega_i}{2} \sigma_z^{(i)} - \hj_x \sum_{i}g_i \sigma_x^{(i)}.
\end{align}
Above $\hj_i$ are the spin operators of the central system, $\sigma_k^{(i)}$ are the Pauli matrices for the $i$-th spin,  $a_i, a_i^\dagger$ are the annihilation and creation operators, respectively, of the environment, 
which is assumed to be thermal with the inverse temperature $\beta$. Both spin-spin and spin-boson models have become of a significant  importance recently due to their role in such fields as e.g. matter-wave interferometry \cite{Hornberger2012, Fein2019}, quantum dots \cite{Urbaszek2013}, nitrogen-vacancy centers \cite{Doherty2013}, applied quantum information \cite{Bergou2021}.

The models fall within our framework with $G=SU(2)$, $H_0=\Omega \hj_z$, $A=-\hj_x$. The adjoint matrix, Eq. \eqref{Ad}, is then already in the canonical form and reads (after rescaling $X_N$):
\begin{align}
R(t)=\left[\begin{matrix} \cos\Omega t & -\sin\Omega t &   \\ \sin\Omega t & \cos\Omega t &  \\ & & 1 \end{matrix}\right].
\end{align}
From Eqs. \eqref{Noddfinal1} and \eqref{Noddfinal2}, we immediately obtain the entropy production (with $a_x=-1$, the rest zero):
\begin{align}
\overline s \approx 2D \left[\Delta \hj_x^2 +\Delta \hj_y^2 +\frac{\gamma}{D}\langle \hj_z\rangle\right],\label{spinJ}
\end{align}
where $D,\gamma$ are from \eqref{Dalpha}, \eqref{gammaalpha} and are given by the well-known expressions \cite{Schlosshauer2007} with $\gamma/D = \tanh(\beta\Omega/2)$.  Eq. \eqref{spinJ}, 
is the same, up to an irrelevant positive prefactor, for both models, which can be seen e.g. using the standard spin-oscillator mapping \cite{Schlosshauer2007}.

We simplify \eqref{spinJ} here by considering  the high temperature limit (cf. \eqref{highT}):
\begin{align}
\overline s \approx 2D \left[ \Delta \hj_x^2 + \Delta \hj_y^2\right].\label{spinJht}
\end{align}
The r.h.s. looks almost like the $G$-invariant dispersion $\langle \Delta C\rangle = \Delta \hj_x^2 + \Delta \hj_y^2 + \Delta \hj_z^2$, which defines the spin coherent states \cite{Perelomov1986}, but without the last term. This absence of the $G$-invariance, changes the minimization problem significantly (Eq. \eqref{spinJht} is invariant only w.r.t. $U(1)$ rotations $e^{i\varphi \hj_z}$). To this end, let us assume that $\hj_i$'s describe a spin-$j$ system. It is then easy to calculate Eq. \eqref{spinJ} for spin coherent states $\ket{\bm n}=e^{i\theta \bm m \bm \hj}\ket{j,-j}$, where $\bm n$ is a unit vector, $\bm m= (\sin\phi, -\cos\phi,0)$, and $\hj_z \ket{j,-j} = -j \ket{j,-j}$. More precisely, we obtain:
\begin{align}
\frac{\overline s}{2D} \approx j\left(1-\frac{1}{2}\sin^2\theta -\frac{\gamma}{D}\cos\theta\right),
\end{align}
with the minimum at $\cos\theta =\gamma/D$ leading to
\begin{align}
\frac{\overline s_{\min}}{2D} \approx \frac{j}{2}\left(1-\frac{\gamma^2}{D^2}\right).
\end{align}
In the special case of $\beta\to 0$, the above equation implies $\overline s_{\min}/2D =j/2$. This is satisfied for spin-$1/2$ systems (e.g. a two-level atom), since then all pure states are coherent states by definition. However, this is not so already for spin-$1$. As we show in Appendix~\ref{append:spin-1}, the minimum for high temperature is achieved for the following $U(1)$ family of states:
\begin{align}\label{psi_j1}
\ket{\psi}=\sqrt{\frac{5}{16}}\left(e^{i\psi}\ket{1,1} +e^{-i\psi}\ket{1,-1}\right) + \sqrt{\frac{3}{8}} \ket{1,0},
\end{align}
and the minimum value is given by
\begin{align}
\frac{\overline{s}_{\min}}{2D} =\frac{7}{16} \approx 0.4375 <0.5,
\end{align} 
which is strictly lower than the minimum for spin coherent states.  Although the overlap $\langle \psi| \bm{n}\rangle$ can be as high as $1/2+\sqrt{15}/8\approx 0.98$ for certain coherent states, the states $\ket\psi$ are superpositions of two orthogonal spin coherent states, cf. \eqref{superpos}, showing their ``non-classicality" w.r.t. spin coherent states. 



\section{Conclusions}
\label{sec:conclusion}
Although the introduction of pointer states  was arguably a serious conceptual step towards understanding of the quantum-to-classical transition and decoherence, there has been no systematic approach of studying them, apart from simplest situations. This work helps to fill this gap.  We develop a general framework for finding pointer states within the Born-Markov approximation and with an additional assumption of a dynamical group structure.  Satisfied in most important open systems models, the latter  allows 
to use group-theoretical methods to find defining equations of pointer states in the weak-coupling limit. We study those equations in detail for spin-$J$ systems with either bosonic or spin environments.

We used mostly compact, semi-simple groups above but as we have mentioned the method is more universal and can be applied to other groups too. It is enough that the free evolution generates a group with some known canonical decomposition, like we showed on the QBM example and the oscillator group. 
The defining equations unfortunately present a difficult optimization problem even in the low-damping regime, as the spin models illustrate, and new mathematical tools will most probably be needed to tackle it. It would be also interesting to learn more about the physical characteristics of the states found here (we know they are not minimum uncertainty states) and understand if and how a general initial state of the system evolves towards their  mixture  (cf. \cite{Eisert2004}). And going one step further,  if advanced quantum-to-classical transition mechanisms such as quantum Darwinism and Spectrum Broadcast Structures  \cite{Zurek2009, Korbicz2014, Le2019, Unden2019, Korbicz2021} can be defined around them. We hope that our work will stimulate a further research into those and related topics and contribute to the debate on the dynamical emergence of classical properties in quantum systems. 

\begin{acknowledgements}
We would like to thank Lorenza Viola for valuable discussions and pointing us to some earlier works on pointer states. U.S. and J.K.K. acknowledge the support by Polish National Science Center (NCN) (Grant No. 2019/35/B/ST2/01896).
\end{acknowledgements}

\bibliography{predictability-sieve_v4}

\begin{thebibliography}{46}%
\makeatletter
\providecommand \@ifxundefined [1]{%
 \@ifx{#1\undefined}
}%
\providecommand \@ifnum [1]{%
 \ifnum #1\expandafter \@firstoftwo
 \else \expandafter \@secondoftwo
 \fi
}%
\providecommand \@ifx [1]{%
 \ifx #1\expandafter \@firstoftwo
 \else \expandafter \@secondoftwo
 \fi
}%
\providecommand \natexlab [1]{#1}%
\providecommand \enquote  [1]{``#1''}%
\providecommand \bibnamefont  [1]{#1}%
\providecommand \bibfnamefont [1]{#1}%
\providecommand \citenamefont [1]{#1}%
\providecommand \href@noop [0]{\@secondoftwo}%
\providecommand \href [0]{\begingroup \@sanitize@url \@href}%
\providecommand \@href[1]{\@@startlink{#1}\@@href}%
\providecommand \@@href[1]{\endgroup#1\@@endlink}%
\providecommand \@sanitize@url [0]{\catcode `\\12\catcode `\$12\catcode
  `\&12\catcode `\#12\catcode `\^12\catcode `\_12\catcode `\%12\relax}%
\providecommand \@@startlink[1]{}%
\providecommand \@@endlink[0]{}%
\providecommand \url  [0]{\begingroup\@sanitize@url \@url }%
\providecommand \@url [1]{\endgroup\@href {#1}{\urlprefix }}%
\providecommand \urlprefix  [0]{URL }%
\providecommand \Eprint [0]{\href }%
\providecommand \doibase [0]{https://doi.org/}%
\providecommand \selectlanguage [0]{\@gobble}%
\providecommand \bibinfo  [0]{\@secondoftwo}%
\providecommand \bibfield  [0]{\@secondoftwo}%
\providecommand \translation [1]{[#1]}%
\providecommand \BibitemOpen [0]{}%
\providecommand \bibitemStop [0]{}%
\providecommand \bibitemNoStop [0]{.\EOS\space}%
\providecommand \EOS [0]{\spacefactor3000\relax}%
\providecommand \BibitemShut  [1]{\csname bibitem#1\endcsname}%
\let\auto@bib@innerbib\@empty
\bibitem [{\citenamefont {Zeh}(1970)}]{Zeh1970}%
  \BibitemOpen
  \bibfield  {author} {\bibinfo {author} {\bibfnamefont {H.~D.}\ \bibnamefont
  {Zeh}},\ }\bibfield  {title} {\bibinfo {title} {On the interpretation of
  measurement in quantum theory},\ }\href
  {https://doi.org/https://doi.org/10.1007/BF00708656} {\bibfield  {journal}
  {\bibinfo  {journal} {Found. Phys.}\ }\textbf {\bibinfo {volume} {1}},\
  \bibinfo {pages} {69} (\bibinfo {year} {1970})}\BibitemShut {NoStop}%
\bibitem [{\citenamefont {Joos}\ \emph {et~al.}(2003)\citenamefont {Joos},
  \citenamefont {Zeh}, \citenamefont {Kiefer}, \citenamefont {Giulini},
  \citenamefont {Kupsch},\ and\ \citenamefont {Stamatescu}}]{Joos2003}%
  \BibitemOpen
  \bibfield  {author} {\bibinfo {author} {\bibfnamefont {E.}~\bibnamefont
  {Joos}}, \bibinfo {author} {\bibfnamefont {H.~D.}\ \bibnamefont {Zeh}},
  \bibinfo {author} {\bibfnamefont {C.}~\bibnamefont {Kiefer}}, \bibinfo
  {author} {\bibfnamefont {D.}~\bibnamefont {Giulini}}, \bibinfo {author}
  {\bibfnamefont {J.}~\bibnamefont {Kupsch}},\ and\ \bibinfo {author}
  {\bibfnamefont {I.-O.}\ \bibnamefont {Stamatescu}},\ }\href
  {https://doi.org/https://doi.org/10.1007/978-3-662-05328-7} {\emph {\bibinfo
  {title} {Decoherence and the Appearance of a Classical World in Quantum
  Theory}}}\ (\bibinfo  {publisher} {Springer Berlin},\ \bibinfo {address}
  {Heidelberg},\ \bibinfo {year} {2003})\BibitemShut {NoStop}%
\bibitem [{\citenamefont {Zurek}(2003)}]{Zurek2003}%
  \BibitemOpen
  \bibfield  {author} {\bibinfo {author} {\bibfnamefont {W.~H.}\ \bibnamefont
  {Zurek}},\ }\bibfield  {title} {\bibinfo {title} {Decoherence, einselection,
  and the quantum origins of the classical},\ }\href
  {https://doi.org/10.1103/RevModPhys.75.715} {\bibfield  {journal} {\bibinfo
  {journal} {Rev. Mod. Phys.}\ }\textbf {\bibinfo {volume} {75}},\ \bibinfo
  {pages} {715} (\bibinfo {year} {2003})}\BibitemShut {NoStop}%
\bibitem [{\citenamefont {Breuer}\ and\ \citenamefont
  {Petruccione}(2002)}]{Breuer2002}%
  \BibitemOpen
  \bibfield  {author} {\bibinfo {author} {\bibfnamefont {H.-P.}\ \bibnamefont
  {Breuer}}\ and\ \bibinfo {author} {\bibfnamefont {F.}~\bibnamefont
  {Petruccione}},\ }\href
  {https://doi.org/10.1093/acprof:oso/9780199213900.001.0001} {\emph {\bibinfo
  {title} {{The Theory of Open Quantum Systems}}}}\ (\bibinfo  {publisher}
  {Oxford University Press},\ \bibinfo {year} {2002})\BibitemShut {NoStop}%
\bibitem [{\citenamefont {Schlosshauer}(2007)}]{Schlosshauer2007}%
  \BibitemOpen
  \bibfield  {author} {\bibinfo {author} {\bibfnamefont {M.}~\bibnamefont
  {Schlosshauer}},\ }\href
  {https://doi.org/https://doi.org/10.1007/978-3-540-35775-9} {\emph {\bibinfo
  {title} {Decoherence and the Quantum-To-Classical Transition}}}\ (\bibinfo
  {publisher} {Springer Berlin},\ \bibinfo {address} {Heidelberg},\ \bibinfo
  {year} {2007})\BibitemShut {NoStop}%
\bibitem [{\citenamefont {Chiorescu}\ \emph {et~al.}(2003)\citenamefont
  {Chiorescu}, \citenamefont {Nakamura}, \citenamefont {Harmans},\ and\
  \citenamefont {Mooij}}]{Chiorescu2003}%
  \BibitemOpen
  \bibfield  {author} {\bibinfo {author} {\bibfnamefont {I.}~\bibnamefont
  {Chiorescu}}, \bibinfo {author} {\bibfnamefont {Y.}~\bibnamefont {Nakamura}},
  \bibinfo {author} {\bibfnamefont {C.~J. P.~M.}\ \bibnamefont {Harmans}},\
  and\ \bibinfo {author} {\bibfnamefont {J.~E.}\ \bibnamefont {Mooij}},\
  }\bibfield  {title} {\bibinfo {title} {Coherent quantum dynamics of a
  superconducting flux qubit},\ }\href
  {https://doi.org/https://doi.org/10.1126/science.1081045} {\bibfield
  {journal} {\bibinfo  {journal} {Science}\ }\textbf {\bibinfo {volume} {21}},\
  \bibinfo {pages} {1869} (\bibinfo {year} {2003})}\BibitemShut {NoStop}%
\bibitem [{\citenamefont {Del\'eglise}\ \emph {et~al.}(2008)\citenamefont
  {Del\'eglise}, \citenamefont {Dotsenko}, \citenamefont {Sayrin},
  \citenamefont {Bernu}, \citenamefont {Brune}, \citenamefont {Raimond},\ and\
  \citenamefont {Haroche}}]{Deleglise2008}%
  \BibitemOpen
  \bibfield  {author} {\bibinfo {author} {\bibfnamefont {S.}~\bibnamefont
  {Del\'eglise}}, \bibinfo {author} {\bibfnamefont {I.}~\bibnamefont
  {Dotsenko}}, \bibinfo {author} {\bibfnamefont {C.}~\bibnamefont {Sayrin}},
  \bibinfo {author} {\bibfnamefont {J.}~\bibnamefont {Bernu}}, \bibinfo
  {author} {\bibfnamefont {M.}~\bibnamefont {Brune}}, \bibinfo {author}
  {\bibfnamefont {J.-M.}\ \bibnamefont {Raimond}},\ and\ \bibinfo {author}
  {\bibfnamefont {S.}~\bibnamefont {Haroche}},\ }\bibfield  {title} {\bibinfo
  {title} {Reconstruction of non-classical cavity field states with snapshots
  of their decoherence.},\ }\href
  {https://doi.org/https://doi.org/10.1038/nature07288} {\bibfield  {journal}
  {\bibinfo  {journal} {Nature}\ }\textbf {\bibinfo {volume} {455}},\ \bibinfo
  {pages} {510} (\bibinfo {year} {2008})}\BibitemShut {NoStop}%
\bibitem [{\citenamefont {H\"affner}\ \emph {et~al.}(2008)\citenamefont
  {H\"affner}, \citenamefont {Roos},\ and\ \citenamefont
  {Blatt}}]{Haffner2008}%
  \BibitemOpen
  \bibfield  {author} {\bibinfo {author} {\bibfnamefont {H.}~\bibnamefont
  {H\"affner}}, \bibinfo {author} {\bibfnamefont {C.}~\bibnamefont {Roos}},\
  and\ \bibinfo {author} {\bibfnamefont {R.}~\bibnamefont {Blatt}},\ }\bibfield
   {title} {\bibinfo {title} {Quantum computing with trapped ions},\ }\href
  {https://doi.org/https://doi.org/10.1016/j.physrep.2008.09.003} {\bibfield
  {journal} {\bibinfo  {journal} {Physics Reports}\ }\textbf {\bibinfo {volume}
  {469}},\ \bibinfo {pages} {155} (\bibinfo {year} {2008})}\BibitemShut
  {NoStop}%
\bibitem [{\citenamefont {Hornberger}\ \emph {et~al.}(2012)\citenamefont
  {Hornberger}, \citenamefont {Gerlich}, \citenamefont {Haslinger},
  \citenamefont {Nimmrichter},\ and\ \citenamefont {Arndt}}]{Hornberger2012}%
  \BibitemOpen
  \bibfield  {author} {\bibinfo {author} {\bibfnamefont {K.}~\bibnamefont
  {Hornberger}}, \bibinfo {author} {\bibfnamefont {S.}~\bibnamefont {Gerlich}},
  \bibinfo {author} {\bibfnamefont {P.}~\bibnamefont {Haslinger}}, \bibinfo
  {author} {\bibfnamefont {S.}~\bibnamefont {Nimmrichter}},\ and\ \bibinfo
  {author} {\bibfnamefont {M.}~\bibnamefont {Arndt}},\ }\bibfield  {title}
  {\bibinfo {title} {Colloquium: Quantum interference of clusters and
  molecules},\ }\href {https://doi.org/10.1103/RevModPhys.84.157} {\bibfield
  {journal} {\bibinfo  {journal} {Rev. Mod. Phys.}\ }\textbf {\bibinfo {volume}
  {84}},\ \bibinfo {pages} {157} (\bibinfo {year} {2012})}\BibitemShut
  {NoStop}%
\bibitem [{\citenamefont {Moser}\ \emph {et~al.}(2014)\citenamefont {Moser},
  \citenamefont {Eichler}, \citenamefont {G\"uttinger}, \citenamefont
  {Dykman},\ and\ \citenamefont {Bachtold}}]{Moser2014}%
  \BibitemOpen
  \bibfield  {author} {\bibinfo {author} {\bibfnamefont {J.}~\bibnamefont
  {Moser}}, \bibinfo {author} {\bibfnamefont {A.}~\bibnamefont {Eichler}},
  \bibinfo {author} {\bibfnamefont {J.}~\bibnamefont {G\"uttinger}}, \bibinfo
  {author} {\bibfnamefont {M.~I.}\ \bibnamefont {Dykman}},\ and\ \bibinfo
  {author} {\bibfnamefont {A.}~\bibnamefont {Bachtold}},\ }\bibfield  {title}
  {\bibinfo {title} {Nanotube mechanical resonators with quality factors of up
  to 5 million},\ }\href
  {https://doi.org/https://doi.org/10.1038/nnano.2014.234} {\bibfield
  {journal} {\bibinfo  {journal} {Nature Nanotech}\ }\textbf {\bibinfo {volume}
  {9}},\ \bibinfo {pages} {1007–1011} (\bibinfo {year} {2014})}\BibitemShut
  {NoStop}%
\bibitem [{\citenamefont {Tighineanu}\ \emph {et~al.}(2018)\citenamefont
  {Tighineanu}, \citenamefont {Dree\ss{}en}, \citenamefont {Flindt},
  \citenamefont {Lodahl},\ and\ \citenamefont {S\o{}rensen}}]{Tighineanu2018}%
  \BibitemOpen
  \bibfield  {author} {\bibinfo {author} {\bibfnamefont {P.}~\bibnamefont
  {Tighineanu}}, \bibinfo {author} {\bibfnamefont {C.~L.}\ \bibnamefont
  {Dree\ss{}en}}, \bibinfo {author} {\bibfnamefont {C.}~\bibnamefont {Flindt}},
  \bibinfo {author} {\bibfnamefont {P.}~\bibnamefont {Lodahl}},\ and\ \bibinfo
  {author} {\bibfnamefont {A.~S.}\ \bibnamefont {S\o{}rensen}},\ }\bibfield
  {title} {\bibinfo {title} {Phonon decoherence of quantum dots in photonic
  structures: Broadening of the zero-phonon line and the role of
  dimensionality},\ }\href {https://doi.org/10.1103/PhysRevLett.120.257401}
  {\bibfield  {journal} {\bibinfo  {journal} {Phys. Rev. Lett.}\ }\textbf
  {\bibinfo {volume} {120}},\ \bibinfo {pages} {257401} (\bibinfo {year}
  {2018})}\BibitemShut {NoStop}%
\bibitem [{\citenamefont {Fein}\ \emph {et~al.}(2019)\citenamefont {Fein},
  \citenamefont {Geyer}, \citenamefont {Zwick}, \citenamefont {Kiałka},
  \citenamefont {Pedalino}, \citenamefont {Mayor}, \citenamefont {Gerlich},\
  and\ \citenamefont {Arndt}}]{Fein2019}%
  \BibitemOpen
  \bibfield  {author} {\bibinfo {author} {\bibfnamefont {Y.~Y.}\ \bibnamefont
  {Fein}}, \bibinfo {author} {\bibfnamefont {P.}~\bibnamefont {Geyer}},
  \bibinfo {author} {\bibfnamefont {P.}~\bibnamefont {Zwick}}, \bibinfo
  {author} {\bibfnamefont {F.}~\bibnamefont {Kiałka}}, \bibinfo {author}
  {\bibfnamefont {S.}~\bibnamefont {Pedalino}}, \bibinfo {author}
  {\bibfnamefont {M.}~\bibnamefont {Mayor}}, \bibinfo {author} {\bibfnamefont
  {S.}~\bibnamefont {Gerlich}},\ and\ \bibinfo {author} {\bibfnamefont
  {M.}~\bibnamefont {Arndt}},\ }\bibfield  {title} {\bibinfo {title} {Quantum
  superposition of molecules beyond 25 kda},\ }\href
  {https://doi.org/https://doi.org/10.1038/s41567-019-0663-9} {\bibfield
  {journal} {\bibinfo  {journal} {Nat. Phys.}\ }\textbf {\bibinfo {volume}
  {15}},\ \bibinfo {pages} {1242–1245} (\bibinfo {year} {2019})}\BibitemShut
  {NoStop}%
\bibitem [{\citenamefont {Zurek}(1981)}]{Zurek1981}%
  \BibitemOpen
  \bibfield  {author} {\bibinfo {author} {\bibfnamefont {W.~H.}\ \bibnamefont
  {Zurek}},\ }\bibfield  {title} {\bibinfo {title} {Pointer basis of quantum
  apparatus: Into what mixture does the wave packet collapse?},\ }\href
  {https://doi.org/10.1103/PhysRevD.24.1516} {\bibfield  {journal} {\bibinfo
  {journal} {Phys. Rev. D}\ }\textbf {\bibinfo {volume} {24}},\ \bibinfo
  {pages} {1516} (\bibinfo {year} {1981})}\BibitemShut {NoStop}%
\bibitem [{\citenamefont {Zurek}(1982)}]{Zurek1982}%
  \BibitemOpen
  \bibfield  {author} {\bibinfo {author} {\bibfnamefont {W.~H.}\ \bibnamefont
  {Zurek}},\ }\bibfield  {title} {\bibinfo {title} {Environment-induced
  superselection rules},\ }\href {https://doi.org/10.1103/PhysRevD.26.1862}
  {\bibfield  {journal} {\bibinfo  {journal} {Phys. Rev. D}\ }\textbf {\bibinfo
  {volume} {26}},\ \bibinfo {pages} {1862} (\bibinfo {year}
  {1982})}\BibitemShut {NoStop}%
\bibitem [{\citenamefont {Zurek}(1993)}]{Zurek1993b}%
  \BibitemOpen
  \bibfield  {author} {\bibinfo {author} {\bibfnamefont {W.~H.}\ \bibnamefont
  {Zurek}},\ }\bibfield  {title} {\bibinfo {title} {{Preferred States,
  Predictability, Classicality and the Environment-Induced Decoherence}},\
  }\href {https://doi.org/10.1143/ptp/89.2.281} {\bibfield  {journal} {\bibinfo
   {journal} {Progress of Theoretical Physics}\ }\textbf {\bibinfo {volume}
  {89}},\ \bibinfo {pages} {281} (\bibinfo {year} {1993})}\BibitemShut
  {NoStop}%
\bibitem [{\citenamefont {Di\'osi}\ and\ \citenamefont
  {Kiefer}(2000)}]{Diosi2000}%
  \BibitemOpen
  \bibfield  {author} {\bibinfo {author} {\bibfnamefont {L.}~\bibnamefont
  {Di\'osi}}\ and\ \bibinfo {author} {\bibfnamefont {C.}~\bibnamefont
  {Kiefer}},\ }\bibfield  {title} {\bibinfo {title} {Robustness and diffusion
  of pointer states},\ }\href {https://doi.org/10.1103/PhysRevLett.85.3552}
  {\bibfield  {journal} {\bibinfo  {journal} {Phys. Rev. Lett.}\ }\textbf
  {\bibinfo {volume} {85}},\ \bibinfo {pages} {3552} (\bibinfo {year}
  {2000})}\BibitemShut {NoStop}%
\bibitem [{\citenamefont {Zurek}\ \emph {et~al.}(1993)\citenamefont {Zurek},
  \citenamefont {Habib},\ and\ \citenamefont {Paz}}]{Zurek1993}%
  \BibitemOpen
  \bibfield  {author} {\bibinfo {author} {\bibfnamefont {W.~H.}\ \bibnamefont
  {Zurek}}, \bibinfo {author} {\bibfnamefont {S.}~\bibnamefont {Habib}},\ and\
  \bibinfo {author} {\bibfnamefont {J.~P.}\ \bibnamefont {Paz}},\ }\bibfield
  {title} {\bibinfo {title} {Coherent states via decoherence},\ }\href
  {https://doi.org/10.1103/PhysRevLett.70.1187} {\bibfield  {journal} {\bibinfo
   {journal} {Phys. Rev. Lett.}\ }\textbf {\bibinfo {volume} {70}},\ \bibinfo
  {pages} {1187} (\bibinfo {year} {1993})}\BibitemShut {NoStop}%
\bibitem [{\citenamefont {Feynman}\ and\ \citenamefont
  {Vernon}(1963)}]{Feynman1963}%
  \BibitemOpen
  \bibfield  {author} {\bibinfo {author} {\bibfnamefont {R.}~\bibnamefont
  {Feynman}}\ and\ \bibinfo {author} {\bibfnamefont {F.~L.}\ \bibnamefont
  {Vernon}},\ }\bibfield  {title} {\bibinfo {title} {The theory of a general
  quantum system interacting with a linear dissipative system},\ }\href
  {https://doi.org/https://doi.org/10.1016/0003-4916(63)90068-X} {\bibfield
  {journal} {\bibinfo  {journal} {Ann. Phys.}\ }\textbf {\bibinfo {volume}
  {24}},\ \bibinfo {pages} {118} (\bibinfo {year} {1963})}\BibitemShut
  {NoStop}%
\bibitem [{\citenamefont {Ullersma}(1966)}]{Ullersma1966}%
  \BibitemOpen
  \bibfield  {author} {\bibinfo {author} {\bibfnamefont {P.}~\bibnamefont
  {Ullersma}},\ }\bibfield  {title} {\bibinfo {title} {An exactly solvable
  model for brownian motion: I. derivation of the langevin equation},\ }\href
  {https://doi.org/https://doi.org/10.1016/0031-8914(66)90105-4} {\bibfield
  {journal} {\bibinfo  {journal} {Physica}\ }\textbf {\bibinfo {volume} {32}},\
  \bibinfo {pages} {27} (\bibinfo {year} {1966})}\BibitemShut {NoStop}%
\bibitem [{\citenamefont {Glauber}(1963)}]{Glauber1963}%
  \BibitemOpen
  \bibfield  {author} {\bibinfo {author} {\bibfnamefont {R.~J.}\ \bibnamefont
  {Glauber}},\ }\bibfield  {title} {\bibinfo {title} {Coherent and incoherent
  states of the radiation field},\ }\href
  {https://doi.org/10.1103/PhysRev.131.2766} {\bibfield  {journal} {\bibinfo
  {journal} {Phys. Rev.}\ }\textbf {\bibinfo {volume} {131}},\ \bibinfo {pages}
  {2766} (\bibinfo {year} {1963})}\BibitemShut {NoStop}%
\bibitem [{\citenamefont {Sudarshan}(1963)}]{Sudarshan1963}%
  \BibitemOpen
  \bibfield  {author} {\bibinfo {author} {\bibfnamefont {E.~C.~G.}\
  \bibnamefont {Sudarshan}},\ }\bibfield  {title} {\bibinfo {title}
  {Equivalence of semiclassical and quantum mechanical descriptions of
  statistical light beams},\ }\href
  {https://doi.org/10.1103/PhysRevLett.10.277} {\bibfield  {journal} {\bibinfo
  {journal} {Phys. Rev. Lett.}\ }\textbf {\bibinfo {volume} {10}},\ \bibinfo
  {pages} {277} (\bibinfo {year} {1963})}\BibitemShut {NoStop}%
\bibitem [{\citenamefont {Eisert}(2004)}]{Eisert2004}%
  \BibitemOpen
  \bibfield  {author} {\bibinfo {author} {\bibfnamefont {J.}~\bibnamefont
  {Eisert}},\ }\bibfield  {title} {\bibinfo {title} {Exact decoherence to
  pointer states in free open quantum systems is universal},\ }\href
  {https://doi.org/10.1103/PhysRevLett.92.210401} {\bibfield  {journal}
  {\bibinfo  {journal} {Phys. Rev. Lett.}\ }\textbf {\bibinfo {volume} {92}},\
  \bibinfo {pages} {210401} (\bibinfo {year} {2004})}\BibitemShut {NoStop}%
\bibitem [{\citenamefont {Perelomov}(1986)}]{Perelomov1986}%
  \BibitemOpen
  \bibfield  {author} {\bibinfo {author} {\bibfnamefont {A.}~\bibnamefont
  {Perelomov}},\ }\href
  {https://doi.org/https://doi.org/10.1007/978-3-642-61629-7} {\emph {\bibinfo
  {title} {Generalized Coherent States and Their Applications}}}\ (\bibinfo
  {publisher} {Springer Berlin},\ \bibinfo {address} {Heidelberg},\ \bibinfo
  {year} {1986})\BibitemShut {NoStop}%
\bibitem [{\citenamefont {Boixo}\ \emph {et~al.}(2007)\citenamefont {Boixo},
  \citenamefont {Viola},\ and\ \citenamefont {Ortiz}}]{Boixo2007}%
  \BibitemOpen
  \bibfield  {author} {\bibinfo {author} {\bibfnamefont {S.}~\bibnamefont
  {Boixo}}, \bibinfo {author} {\bibfnamefont {L.}~\bibnamefont {Viola}},\ and\
  \bibinfo {author} {\bibfnamefont {G.}~\bibnamefont {Ortiz}},\ }\bibfield
  {title} {\bibinfo {title} {Generalized coherent states as preferred states of
  open quantum systems},\ }\href {https://doi.org/10.1209/0295-5075/79/40003}
  {\bibfield  {journal} {\bibinfo  {journal} {Europhysics Letters}\ }\textbf
  {\bibinfo {volume} {79}},\ \bibinfo {pages} {40003} (\bibinfo {year}
  {2007})}\BibitemShut {NoStop}%
\bibitem [{\citenamefont {Kossakowski}(1972)}]{Kossakowski1972}%
  \BibitemOpen
  \bibfield  {author} {\bibinfo {author} {\bibfnamefont {A.}~\bibnamefont
  {Kossakowski}},\ }\bibfield  {title} {\bibinfo {title} {On quantum
  statistical mechanics of non-hamiltonian systems},\ }\href
  {https://doi.org/https://doi.org/10.1016/0034-4877(72)90010-9} {\bibfield
  {journal} {\bibinfo  {journal} {Reports on Mathematical Physics}\ }\textbf
  {\bibinfo {volume} {3}},\ \bibinfo {pages} {247} (\bibinfo {year}
  {1972})}\BibitemShut {NoStop}%
\bibitem [{\citenamefont {Davies}(1974)}]{Davies1974}%
  \BibitemOpen
  \bibfield  {author} {\bibinfo {author} {\bibfnamefont {E.~B.}\ \bibnamefont
  {Davies}},\ }\bibfield  {title} {\bibinfo {title} {Markovian master
  equations},\ }\href {https://doi.org/10.1007/BF01608389} {\bibfield
  {journal} {\bibinfo  {journal} {Commun. Math. Phys.}\ }\textbf {\bibinfo
  {volume} {39}},\ \bibinfo {pages} {91} (\bibinfo {year} {1974})}\BibitemShut
  {NoStop}%
\bibitem [{\citenamefont {Gorini}\ \emph {et~al.}(1976)\citenamefont {Gorini},
  \citenamefont {Kossakowski},\ and\ \citenamefont {Sudarshan}}]{Gorini1976}%
  \BibitemOpen
  \bibfield  {author} {\bibinfo {author} {\bibfnamefont {V.}~\bibnamefont
  {Gorini}}, \bibinfo {author} {\bibfnamefont {A.}~\bibnamefont
  {Kossakowski}},\ and\ \bibinfo {author} {\bibfnamefont {E.~C.~G.}\
  \bibnamefont {Sudarshan}},\ }\bibfield  {title} {\bibinfo {title} {Completely
  positive dynamical semigroups of n‐level systems},\ }\href
  {https://doi.org/10.1063/1.522979} {\bibfield  {journal} {\bibinfo  {journal}
  {J. Math. Phys.}\ }\textbf {\bibinfo {volume} {17}},\ \bibinfo {pages} {821}
  (\bibinfo {year} {1976})}\BibitemShut {NoStop}%
\bibitem [{\citenamefont {Lindblad}(1976)}]{Lindblad1976}%
  \BibitemOpen
  \bibfield  {author} {\bibinfo {author} {\bibfnamefont {G.}~\bibnamefont
  {Lindblad}},\ }\bibfield  {title} {\bibinfo {title} {On the generators of
  quantum dynamical semigroups},\ }\href {https://doi.org/10.1007/BF01608499}
  {\bibfield  {journal} {\bibinfo  {journal} {Commun. Math. Phys.}\ }\textbf
  {\bibinfo {volume} {48}},\ \bibinfo {pages} {119} (\bibinfo {year}
  {1976})}\BibitemShut {NoStop}%
\bibitem [{\citenamefont {Paraoanu}\ and\ \citenamefont
  {Scutaru}(1998)}]{Paraoanu1998}%
  \BibitemOpen
  \bibfield  {author} {\bibinfo {author} {\bibfnamefont {G.-S.}\ \bibnamefont
  {Paraoanu}}\ and\ \bibinfo {author} {\bibfnamefont {H.}~\bibnamefont
  {Scutaru}},\ }\bibfield  {title} {\bibinfo {title} {Classical states via
  decoherence},\ }\href
  {https://doi.org/https://doi.org/10.1016/S0375-9601(97)00925-0} {\bibfield
  {journal} {\bibinfo  {journal} {Phys. Lett. A}\ }\textbf {\bibinfo {volume}
  {238}},\ \bibinfo {pages} {219} (\bibinfo {year} {1998})}\BibitemShut
  {NoStop}%
\bibitem [{\citenamefont {Khodjasteh}\ \emph {et~al.}(2011)\citenamefont
  {Khodjasteh}, \citenamefont {Dobrovitski},\ and\ \citenamefont
  {Viola}}]{Khodjasteh2011}%
  \BibitemOpen
  \bibfield  {author} {\bibinfo {author} {\bibfnamefont {K.}~\bibnamefont
  {Khodjasteh}}, \bibinfo {author} {\bibfnamefont {V.~V.}\ \bibnamefont
  {Dobrovitski}},\ and\ \bibinfo {author} {\bibfnamefont {L.}~\bibnamefont
  {Viola}},\ }\bibfield  {title} {\bibinfo {title} {Pointer states via
  engineered dissipation},\ }\href {https://doi.org/10.1103/PhysRevA.84.022336}
  {\bibfield  {journal} {\bibinfo  {journal} {Phys. Rev. A}\ }\textbf {\bibinfo
  {volume} {84}},\ \bibinfo {pages} {022336} (\bibinfo {year}
  {2011})}\BibitemShut {NoStop}%
\bibitem [{\citenamefont {Donker}\ \emph {et~al.}(2017)\citenamefont {Donker},
  \citenamefont {De~Raedt},\ and\ \citenamefont {Katsnelson}}]{Donker2017}%
  \BibitemOpen
  \bibfield  {author} {\bibinfo {author} {\bibfnamefont {H.~C.}\ \bibnamefont
  {Donker}}, \bibinfo {author} {\bibfnamefont {H.}~\bibnamefont {De~Raedt}},\
  and\ \bibinfo {author} {\bibfnamefont {M.~I.}\ \bibnamefont {Katsnelson}},\
  }\bibfield  {title} {\bibinfo {title} {{Decoherence and pointer states in
  small antiferromagnets: A benchmark test}},\ }\href
  {https://doi.org/10.21468/SciPostPhys.2.2.010} {\bibfield  {journal}
  {\bibinfo  {journal} {SciPost Phys.}\ }\textbf {\bibinfo {volume} {2}},\
  \bibinfo {pages} {010} (\bibinfo {year} {2017})}\BibitemShut {NoStop}%
\bibitem [{\citenamefont {Winczewski}\ \emph {et~al.}(2021)\citenamefont
  {Winczewski}, \citenamefont {Mandarino}, \citenamefont {Horodecki},\ and\
  \citenamefont {Alicki}}]{Winczewski2021}%
  \BibitemOpen
  \bibfield  {author} {\bibinfo {author} {\bibfnamefont {M.}~\bibnamefont
  {Winczewski}}, \bibinfo {author} {\bibfnamefont {A.}~\bibnamefont
  {Mandarino}}, \bibinfo {author} {\bibfnamefont {M.}~\bibnamefont
  {Horodecki}},\ and\ \bibinfo {author} {\bibfnamefont {R.}~\bibnamefont
  {Alicki}},\ }\href {https://arxiv.org/abs/2106.05776} {\bibinfo {title}
  {Bypassing the intermediate times dilemma for open quantum system}} (\bibinfo
  {year} {2021}),\ \Eprint {https://arxiv.org/abs/2106.05776} {arXiv:2106.05776
  [quant-ph]} \BibitemShut {NoStop}%
\bibitem [{\citenamefont {Tupkary}\ \emph {et~al.}(2022)\citenamefont
  {Tupkary}, \citenamefont {Dhar}, \citenamefont {Kulkarni},\ and\
  \citenamefont {Purkayastha}}]{Tupkary2022}%
  \BibitemOpen
  \bibfield  {author} {\bibinfo {author} {\bibfnamefont {D.}~\bibnamefont
  {Tupkary}}, \bibinfo {author} {\bibfnamefont {A.}~\bibnamefont {Dhar}},
  \bibinfo {author} {\bibfnamefont {M.}~\bibnamefont {Kulkarni}},\ and\
  \bibinfo {author} {\bibfnamefont {A.}~\bibnamefont {Purkayastha}},\
  }\bibfield  {title} {\bibinfo {title} {Fundamental limitations in lindblad
  descriptions of systems weakly coupled to baths},\ }\href
  {https://doi.org/10.1103/PhysRevA.105.032208} {\bibfield  {journal} {\bibinfo
   {journal} {Phys. Rev. A}\ }\textbf {\bibinfo {volume} {105}},\ \bibinfo
  {pages} {032208} (\bibinfo {year} {2022})}\BibitemShut {NoStop}%
\bibitem [{\citenamefont {Redfield}(1957)}]{Redfield1957}%
  \BibitemOpen
  \bibfield  {author} {\bibinfo {author} {\bibfnamefont {A.~G.}\ \bibnamefont
  {Redfield}},\ }\bibfield  {title} {\bibinfo {title} {On the theory of
  relaxation processes},\ }\href {https://doi.org/10.1147/rd.11.0019}
  {\bibfield  {journal} {\bibinfo  {journal} {IBM Journal of Research and
  Development}\ }\textbf {\bibinfo {volume} {1}},\ \bibinfo {pages} {19}
  (\bibinfo {year} {1957})}\BibitemShut {NoStop}%
\bibitem [{\citenamefont {Bloch}(1957)}]{Bloch1957}%
  \BibitemOpen
  \bibfield  {author} {\bibinfo {author} {\bibfnamefont {F.}~\bibnamefont
  {Bloch}},\ }\bibfield  {title} {\bibinfo {title} {Generalized theory of
  relaxation},\ }\href {https://doi.org/10.1103/PhysRev.105.1206} {\bibfield
  {journal} {\bibinfo  {journal} {Phys. Rev.}\ }\textbf {\bibinfo {volume}
  {105}},\ \bibinfo {pages} {1206} (\bibinfo {year} {1957})}\BibitemShut
  {NoStop}%
\bibitem [{\citenamefont {Kirillov~Jr.}(2008)}]{kirillov2008}%
  \BibitemOpen
  \bibfield  {author} {\bibinfo {author} {\bibfnamefont {A.}~\bibnamefont
  {Kirillov~Jr.}},\ }\href {https://doi.org/10.1017/CBO9780511755156} {\emph
  {\bibinfo {title} {An Introduction to Lie Groups and Lie Algebras}}},\
  Cambridge Studies in Advanced Mathematics\ (\bibinfo  {publisher} {Cambridge
  University Press},\ \bibinfo {year} {2008})\BibitemShut {NoStop}%
\bibitem [{\citenamefont {Streater}(1967)}]{Streater1967}%
  \BibitemOpen
  \bibfield  {author} {\bibinfo {author} {\bibfnamefont {R.~F.}\ \bibnamefont
  {Streater}},\ }\bibfield  {title} {\bibinfo {title} {The representations of
  the oscillator group},\ }\href {https://doi.org/10.1007/BF01645431}
  {\bibfield  {journal} {\bibinfo  {journal} {Commun. Math. Phys.}\ }\textbf
  {\bibinfo {volume} {4}},\ \bibinfo {pages} {217} (\bibinfo {year}
  {1967})}\BibitemShut {NoStop}%
\bibitem [{\citenamefont {Groenewold}(1946)}]{Groenewold1946}%
  \BibitemOpen
  \bibfield  {author} {\bibinfo {author} {\bibfnamefont {H.}~\bibnamefont
  {Groenewold}},\ }\bibfield  {title} {\bibinfo {title} {On the principles of
  elementary quantum mechanics},\ }\href
  {https://doi.org/https://doi.org/10.1016/S0031-8914(46)80059-4} {\bibfield
  {journal} {\bibinfo  {journal} {Physica}\ }\textbf {\bibinfo {volume} {12}},\
  \bibinfo {pages} {405} (\bibinfo {year} {1946})}\BibitemShut {NoStop}%
\bibitem [{\citenamefont {Urbaszek}\ \emph {et~al.}(2013)\citenamefont
  {Urbaszek}, \citenamefont {Marie}, \citenamefont {Amand}, \citenamefont
  {Krebs}, \citenamefont {Voisin}, \citenamefont {Maletinsky}, \citenamefont
  {H\"ogele},\ and\ \citenamefont {Imamoglu}}]{Urbaszek2013}%
  \BibitemOpen
  \bibfield  {author} {\bibinfo {author} {\bibfnamefont {B.}~\bibnamefont
  {Urbaszek}}, \bibinfo {author} {\bibfnamefont {X.}~\bibnamefont {Marie}},
  \bibinfo {author} {\bibfnamefont {T.}~\bibnamefont {Amand}}, \bibinfo
  {author} {\bibfnamefont {O.}~\bibnamefont {Krebs}}, \bibinfo {author}
  {\bibfnamefont {P.}~\bibnamefont {Voisin}}, \bibinfo {author} {\bibfnamefont
  {P.}~\bibnamefont {Maletinsky}}, \bibinfo {author} {\bibfnamefont
  {A.}~\bibnamefont {H\"ogele}},\ and\ \bibinfo {author} {\bibfnamefont
  {A.}~\bibnamefont {Imamoglu}},\ }\bibfield  {title} {\bibinfo {title}
  {Nuclear spin physics in quantum dots: An optical investigation},\ }\href
  {https://doi.org/10.1103/RevModPhys.85.79} {\bibfield  {journal} {\bibinfo
  {journal} {Rev. Mod. Phys.}\ }\textbf {\bibinfo {volume} {85}},\ \bibinfo
  {pages} {79} (\bibinfo {year} {2013})}\BibitemShut {NoStop}%
\bibitem [{\citenamefont {Doherty}\ \emph {et~al.}(2013)\citenamefont
  {Doherty}, \citenamefont {Manson}, \citenamefont {Delaney}, \citenamefont
  {Jelezko}, \citenamefont {Wrachtrup},\ and\ \citenamefont
  {Hollenberg}}]{Doherty2013}%
  \BibitemOpen
  \bibfield  {author} {\bibinfo {author} {\bibfnamefont {M.~W.}\ \bibnamefont
  {Doherty}}, \bibinfo {author} {\bibfnamefont {N.~B.}\ \bibnamefont {Manson}},
  \bibinfo {author} {\bibfnamefont {P.}~\bibnamefont {Delaney}}, \bibinfo
  {author} {\bibfnamefont {F.}~\bibnamefont {Jelezko}}, \bibinfo {author}
  {\bibfnamefont {J.}~\bibnamefont {Wrachtrup}},\ and\ \bibinfo {author}
  {\bibfnamefont {L.~C.}\ \bibnamefont {Hollenberg}},\ }\bibfield  {title}
  {\bibinfo {title} {The nitrogen-vacancy colour centre in diamond},\ }\href
  {https://doi.org/https://doi.org/10.1016/j.physrep.2013.02.001} {\bibfield
  {journal} {\bibinfo  {journal} {Physics Reports}\ }\textbf {\bibinfo {volume}
  {528}},\ \bibinfo {pages} {1} (\bibinfo {year} {2013})},\ \bibinfo {note}
  {the nitrogen-vacancy colour centre in diamond}\BibitemShut {NoStop}%
\bibitem [{\citenamefont {Bergou}\ \emph {et~al.}(2021)\citenamefont {Bergou},
  \citenamefont {Hillery},\ and\ \citenamefont {Saffman}}]{Bergou2021}%
  \BibitemOpen
  \bibfield  {author} {\bibinfo {author} {\bibfnamefont {J.~A.}\ \bibnamefont
  {Bergou}}, \bibinfo {author} {\bibfnamefont {M.}~\bibnamefont {Hillery}},\
  and\ \bibinfo {author} {\bibfnamefont {M.}~\bibnamefont {Saffman}},\ }\href
  {https://doi.org/10.1007/978-3-030-75436-5_9} {\emph {\bibinfo {title}
  {Quantum Information Processing: Theory and Implementation}}}\ (\bibinfo
  {publisher} {Springer International Publishing},\ \bibinfo {address} {Cham},\
  \bibinfo {year} {2021})\BibitemShut {NoStop}%
\bibitem [{\citenamefont {Zurek}(2009)}]{Zurek2009}%
  \BibitemOpen
  \bibfield  {author} {\bibinfo {author} {\bibfnamefont {W.~H.}\ \bibnamefont
  {Zurek}},\ }\bibfield  {title} {\bibinfo {title} {Quantum darwinism},\ }\href
  {https://doi.org/10.1038/nphys1202} {\bibfield  {journal} {\bibinfo
  {journal} {Nat. Phys.}\ }\textbf {\bibinfo {volume} {5}},\ \bibinfo {pages}
  {181} (\bibinfo {year} {2009})}\BibitemShut {NoStop}%
\bibitem [{\citenamefont {Korbicz}\ \emph {et~al.}(2014)\citenamefont
  {Korbicz}, \citenamefont {Horodecki},\ and\ \citenamefont
  {Horodecki}}]{Korbicz2014}%
  \BibitemOpen
  \bibfield  {author} {\bibinfo {author} {\bibfnamefont {J.~K.}\ \bibnamefont
  {Korbicz}}, \bibinfo {author} {\bibfnamefont {P.}~\bibnamefont {Horodecki}},\
  and\ \bibinfo {author} {\bibfnamefont {R.}~\bibnamefont {Horodecki}},\
  }\bibfield  {title} {\bibinfo {title} {Objectivity in a noisy photonic
  environment through quantum state information broadcasting},\ }\href
  {https://doi.org/10.1103/PhysRevLett.112.120402} {\bibfield  {journal}
  {\bibinfo  {journal} {Phys. Rev. Lett.}\ }\textbf {\bibinfo {volume} {112}},\
  \bibinfo {pages} {120402} (\bibinfo {year} {2014})}\BibitemShut {NoStop}%
\bibitem [{\citenamefont {Le}\ and\ \citenamefont
  {Olaya-Castro}(2019)}]{Le2019}%
  \BibitemOpen
  \bibfield  {author} {\bibinfo {author} {\bibfnamefont {T.~P.}\ \bibnamefont
  {Le}}\ and\ \bibinfo {author} {\bibfnamefont {A.}~\bibnamefont
  {Olaya-Castro}},\ }\bibfield  {title} {\bibinfo {title} {Strong quantum
  darwinism and strong independence are equivalent to spectrum broadcast
  structure},\ }\href {https://doi.org/10.1103/PhysRevLett.122.010403}
  {\bibfield  {journal} {\bibinfo  {journal} {Phys. Rev. Lett.}\ }\textbf
  {\bibinfo {volume} {122}},\ \bibinfo {pages} {010403} (\bibinfo {year}
  {2019})}\BibitemShut {NoStop}%
\bibitem [{\citenamefont {Unden}\ \emph {et~al.}(2019)\citenamefont {Unden},
  \citenamefont {Louzon}, \citenamefont {Zwolak}, \citenamefont {Zurek},\ and\
  \citenamefont {Jelezko}}]{Unden2019}%
  \BibitemOpen
  \bibfield  {author} {\bibinfo {author} {\bibfnamefont {T.~K.}\ \bibnamefont
  {Unden}}, \bibinfo {author} {\bibfnamefont {D.}~\bibnamefont {Louzon}},
  \bibinfo {author} {\bibfnamefont {M.}~\bibnamefont {Zwolak}}, \bibinfo
  {author} {\bibfnamefont {W.~H.}\ \bibnamefont {Zurek}},\ and\ \bibinfo
  {author} {\bibfnamefont {F.}~\bibnamefont {Jelezko}},\ }\bibfield  {title}
  {\bibinfo {title} {Revealing the emergence of classicality using
  nitrogen-vacancy centers},\ }\href
  {https://doi.org/10.1103/PhysRevLett.123.140402} {\bibfield  {journal}
  {\bibinfo  {journal} {Phys. Rev. Lett.}\ }\textbf {\bibinfo {volume} {123}},\
  \bibinfo {pages} {140402} (\bibinfo {year} {2019})}\BibitemShut {NoStop}%
\bibitem [{\citenamefont {Korbicz}(2021)}]{Korbicz2021}%
  \BibitemOpen
  \bibfield  {author} {\bibinfo {author} {\bibfnamefont {J.~K.}\ \bibnamefont
  {Korbicz}},\ }\bibfield  {title} {\bibinfo {title} {Roads to objectivity:
  {Q}uantum {D}arwinism, {S}pectrum {B}roadcast {S}tructures, and {S}trong
  quantum {D}arwinism--a review},\ }\href
  {https://doi.org/10.22331/q-2021-11-08-571} {\bibfield  {journal} {\bibinfo
  {journal} {{Quantum}}\ }\textbf {\bibinfo {volume} {5}},\ \bibinfo {pages}
  {571} (\bibinfo {year} {2021})}\BibitemShut {NoStop}%
\end{thebibliography}%

\appendix

\section{Spin-\texorpdfstring{$1$}{1} systems}\label{append:spin-1}
Here we minimize analytically the entropy production for spin-$1$ systems. In this case $J_i$ can be written as
\begin{align}
&\hj_x=\frac{1}{\sqrt{2}}\begin{pmatrix}
0 & 1 & 0\\ 1 & 0 & 1\\0 & 1 &0
\end{pmatrix}; ~~\hj_y=\frac{1}{\sqrt{2}}\begin{pmatrix}
0 & -i & 0\\ i & 0 & -i\\0 & i &0
\end{pmatrix}; \\& \hj_z=\begin{pmatrix}
1 & 0 & 0\\ 0 & 0 & 0\\0 & 0&-1
\end{pmatrix}.
\end{align}
Let the initial state $\ket\psi$ be parametrized using the usual eigenstates of $J^2,J_z$ as follows:
\begin{align}
\ket\psi = q\ket{1,1}+s\ket{1,-1}+r\ket{1,0}\label{psiparam}.
\end{align}
We then have
\begin{align}
\label{eq:jx-average}
&  \hj_z =\bra{\psi}\hj_z\ket{\psi} =|q|^2-|s|^2,\\
\label{eq:delta-jz}
& \Delta \hj_x^2 =\frac{|q+s|^2}{2}+r^2 -2 r^2\left(\mathrm{Re}[q+s]\right)^2,\\
\label{eq:delta-jy}
& \Delta \hj_y^2 =\frac{|q-s|^2}{2}+r^2 -2 r^2\left(\mathrm{Im}[q-s]\right)^2.
\end{align}

Taking $q=u+iv$ and $s=\tilde{u}+i\tilde{v}$, the Eq. \eqref{spinJ} for spin-$1$ particles reads
\begin{align}
\label{eq:accu-mix2}
\frac{\overline{s}}{2D }&=1+r^2 -2r^2\left[(u+\tu)^2+(v-\tilde{v})^2\right]\nonumber\\
& +\frac{\gamma}{D}(u^2+v^2-\tilde{u}^2-\tilde{v}^2).
\end{align}
We want to minimize above quantity as a function of $u,v,\tu,\tv$, and $r$. To this end define a Lagrangian function
\begin{widetext}
\begin{align}
\label{eq:f-value}
\mathcal{L}(u,v,\tu,\tv,r,\mu)&=1+r^2 -2r^2\left[(u+\tu)^2+(v-\tilde{v})^2\right] +\frac{\gamma}{D}(u^2+v^2-\tilde{u}^2-\tilde{v}^2)+\mu\left[u^2+v^2+\tilde{u}^2+\tilde{v}^2+r^2-1\right],
\end{align}
where $\mu\geq 0$ is a Lagrange multiplier. To obtain the minima, we set first order derivatives of $\mc{L}$ wrt $u,v,\tu,\tv,r$, and $\mu$ to zero. We have following: 
\begin{subequations}
\begin{align}
&\frac{\partial \mc{L}}{\partial u}=2\left[-2r^2(u+\tu)+\left(\frac{\gamma}{D}+\mu\right) u\right]=0\\
&\frac{\partial \mc{L}}{\partial \tu}=2\left[-2r^2(u+\tu)-\left(\frac{\gamma}{D}-\mu\right) \tu\right]=0\\
&\frac{\partial \mc{L}}{\partial v}=2\left[-2r^2(v-\tilde{v})+\left(\frac{\gamma}{D}+\mu\right) v\right]=0\\
&\frac{\partial \mc{L}}{\partial \tv}=2\left[2r^2(v-\tilde{v})-\left(\frac{\gamma}{D}-\mu\right) \tilde{v}\right]=0\\
&\frac{\partial \mc{L}}{\partial r}=2r\left[(\mu+1) - 2 ((u+\tu)^2+(v-\tilde{v})^2) \right]=0\\
&\frac{\partial \mc{L}}{\partial \mu}=u^2+v^2+\tilde{u}^2+\tilde{v}^2+r^2-1=0.
\end{align}
\end{subequations}
For $r\neq 0$, we have following set of equations equivalent to above equations. 
\begin{subequations}
\begin{align}
&2r^2=\left(\frac{\gamma}{D}+\mu\right) \frac{u}{u+\tu} =- \left(\frac{\gamma}{D}-\mu\right) \frac{\tu}{u+\tu}
=\left(\frac{\gamma}{D}+\mu\right)\frac{ v}{v-\tilde{v}} = \left(\frac{\gamma}{D}-\mu\right)\frac{ \tv}{v-\tilde{v}}\\
&(\mu+1) = 2 ((u+\tu)^2+(v-\tilde{v})^2)\\
&u^2+v^2+\tilde{u}^2+\tilde{v}^2+r^2=1.
\end{align}
\end{subequations}
\end{widetext}
Taking $k=\frac{\frac{\gamma}{D}+\mu}{\frac{\gamma}{D}-\mu}$, we have $\tu=-k u$ and $\tv=k v$, then the above equations become 
\begin{subequations}
\begin{align}
&-2r^2=\frac{\left(\frac{\gamma}{D}\right)^2 -\mu^2}{2\mu};\\
&(\mu+1) = \frac{4\mu^2}{\left(\frac{\gamma}{D}\right)^2+\mu^2} (1-r^2);\\
&u^2 + v^2 =\frac{1}{2} \frac{\left(\frac{\gamma}{D}-\mu\right)^2}{\left(\frac{\gamma}{D}\right)^2+\mu^2} (1-r^2).
\end{align}
\end{subequations}
In the case when $\gamma/D=\mu$, we can define $k'=1/k$, $u=-k'\tu$ and $v=k'\tv$. Combining above equations, we get the following cubic polynomial in terms of $\mu$:
\begin{align}
 2\mu^3 -3 \mu^2+ \left(\frac{\gamma}{D}\right)^2 =0.
\end{align}
Let $\mu=\mu_0$ be a real solution to the above equation giving rise to the minima. Then the pure state achieving the minima in Eq. \eqref{eq:accu-mix2} is give by following parameters: $u, v$ are arbitrary real numbers and 
\begin{align}
\label{eq:mimimum-parameter}
&r=\sqrt{\frac{\mu_0^2-\left(\frac{\gamma}{D}\right)^2}{4\mu_0}};
u^2 + v^2 =\frac{1}{2} \frac{\left(\frac{\gamma}{D}-\mu_0\right)^2}{\left(\frac{\gamma}{D}\right)^2+\mu_0^2} (1-r^2).
\end{align}
The minimum value for the Eq. \eqref{eq:accu-mix2} is given by
\begin{align}
\label{eq:accu-mix3}
\frac{\overline{s}_{\min}}{2D} =\frac{1}{4\mu_0}\left[\mu_0^3-3\mu_0^2 + \left(4-\frac{\gamma^2}{D^2}\right)\mu_0-\frac{\gamma^2}{D^2}\right].
\end{align}

\subsection{ High temperature regime \texorpdfstring{$\frac{\gamma}{D}\approx 0$}{gdH}}
In this case, $\mu_0$ is a solution of the equation $2\mu^3 -3 \mu^2 =0$. Then the minimum of Eq. \eqref{eq:accu-mix2} is obtained for $\mu_0=3/2$ and the pure state achieving the minimum is given by
\begin{align}
\ket{\psi_0}=q\ket{1,1} + \sqrt{\frac{3}{8}} \ket{1,0} + q^*\ket{1,-1},
\end{align}
where $|q|^2= 5/16$. The minimum value of Eq. \eqref{eq:accu-mix2} is given by
\begin{align}
\label{eq:accu-mix4}
\frac{\overline{s}_{\min}}{2D}=\frac{7}{16} = 0.4375.
\end{align}
Figure \ref{fig:scatter-plot} shows that indeed $0.4375$ is the minimum value of $\overline{s}/(2D)$. 
The random pure states on the plot are generated by first sampling two random complex numbers and one random real number using Mathematica functions $\mathsf{RandomComplex[]}$ and $\mathsf{RandomReal[]}$, respectively and then normalizing these three random numbers (cf. Eq. \eqref{psiparam}).
Further, for the state achieving the minima, we have
\begin{align}
&\langle \hj_z\rangle =0,~ \Delta \hj_x^2 =\frac{3}{8}-\mathrm{Re}[q]^2,~ \Delta \hj_y^2  =\frac{3}{8}-\mathrm{Im}[q]^2.
\end{align}
Their overlap with spin-$1$ coherent states is given by
\begin{align}
\langle \psi| \bm{n}\rangle &= \sqrt{\frac{5}{16}}e^{i\varphi}\left(\cos^2\frac{\theta}{2}+e^{-2i(\phi+\varphi)}\sin^2\frac{\theta}{2}\right)\nonumber\\
&-\frac{\sqrt 3}{4}e^{-i\phi}\sin\theta,
\end{align}
where the maximum overlap is with the coherent state, given by
\begin{align}
& \ket{\theta=\frac{\pi}{2},\phi=\pi-\psi} = \nonumber \\
& \frac{1}{2} \ket{1,-1} +  \frac{e^{i\psi}}{\sqrt 2} \ket{1,0} + \frac{e^{2 i\psi}}{ 2} \ket{1,1} 
\end{align}
and reads $(\sqrt 5 + \sqrt 3)^2/16 \approx 0.98$. In the above, we have identified the coherent state $\ket{{\bf n}(\theta, \phi)}$ with $\ket{\theta,\phi}$. Further, decomposing $\ket \psi$ into spin-coherent states shows that it can be represented as a sum of two orthogonal coherent states:
\begin{align}\label{superpos}
    \ket\psi 
=\frac{\sqrt{5}-\sqrt{3}}{4}\ket{\frac{\pi}{2},-\psi} +
\frac{\sqrt{5}+\sqrt{3}}{4} \ket{\frac{\pi}{2},\pi-\psi}.
\end{align}

\begin{figure}
\centering
\includegraphics[width=80mm]{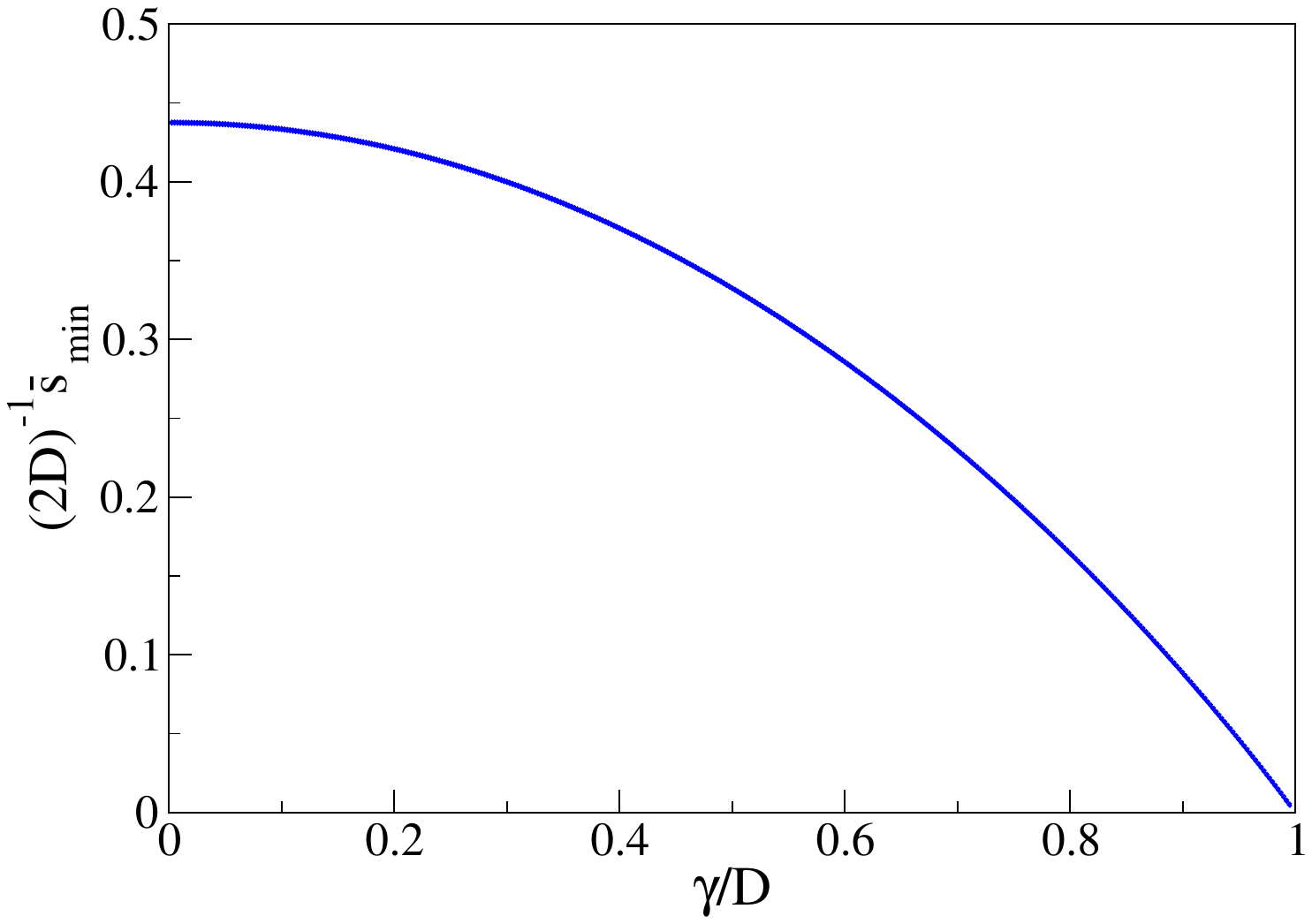}
\caption{The plot shows value of the function $\overline{s}_{\min}/2D$ for a spin-$1$ system, for various values of  $\gamma/D$ that are decided by the temperature of environment. $\gamma/D$ varies from $0$ to $1$ as we decrease temperature. }
\label{fig:temp-vs-mixedness}
\end{figure}

\subsection{Intemediate temperature regime, \texorpdfstring{$\frac{\gamma}{D}=\frac{1}{\sqrt{2}}$}{gdI}}
In this case, $\mu_0$ is a solution of the equation $2\mu^3 -3 \mu^2+\frac{1}{2} =0$. The minimum of Eq. \eqref{eq:accu-mix2} is obtained for $\mu_0=\frac{1+\sqrt{3}}{2}$. Then, the pure state achieving the minimum (that is the pointer state) is given by
\begin{align}
\ket{\psi_0}=q\ket{1,1} + \frac{1}{2} \ket{1,0} -k q^*\ket{1,-1},
\end{align}
where $|q|^2(1+k^2)=\frac{3}{4}$ and
\begin{align}
k=-\frac{1+\sqrt{3} +\sqrt{2}}{1+\sqrt{3}-\sqrt{2}}.
\end{align}

Then using Eq. \eqref{eq:accu-mix3}, the minimum value of Eq. \eqref{eq:accu-mix2} is given by
\begin{align}
\label{eq:accu-mix4-1}
\frac{\overline{s}_{\min}}{2D}
&=\frac{7-3\sqrt{3}}{8}\approx 0.225481.
\end{align}
Further, for the state achieving the minima, we have
\begin{align}
&\langle \hj_z\rangle =-\sqrt{\frac{2}{3}};\nonumber\\
&\Delta\hj_x^2 =\frac{1}{2}\left[1-(1-k)^2 \left(\mathrm{Re}[q]\right)^2+(1+k)^2 \left(\mathrm{Im}[q]\right)^2\right];\nonumber\\
&\Delta\hj_y^2=\frac{1}{2}\left[1+(1+k)^2 \left(\mathrm{Re}[q]\right)^2-(1-k)^2 \left(\mathrm{Im}[q]\right)^2\right].\nonumber
\end{align}

\subsubsection{Low temperature regime, \texorpdfstring{$\frac{\gamma}{D} \approx 1$}{gdL}}
In this case, $\mu_0$ is a solution of the equation $2\mu^3 -3 \mu^2+1 =0$. The minimum of Eq. \eqref{eq:accu-mix2} is obtained for $\mu_0=1$. Then, the pure state achieving the minimum (that is the pointer state) is given by
\begin{align}
\ket{\psi_0}=q\ket{1,-1},
\end{align}
where $|q|^2=1$ or simply $\ket{1,-1}$. The minimum value of Eq. \eqref{eq:accu-mix2} is given by
\begin{align}
\label{eq:accu-mix4-2}
\frac{\overline{s}_{\min}}{2D}&=0.
\end{align}
Further, for the state achieving the minima, we have
\begin{align}
&\langle \hj_z\rangle =-1;~\Delta\hj_x^2 =\frac{1}{2}=\Delta\hj_y^2.
\end{align}

Figure \ref{fig:temp-vs-mixedness} shows that the time averaged mixedness decreases as we decrease temperature.

\end{document}